\documentclass[10pt]{article}

\usepackage{jheppub} 

\usepackage[utf8]{inputenc}
\usepackage[english]{babel}
\usepackage{microtype}
\usepackage{tikzsymbols}

\usepackage{amsmath,amssymb,amsfonts,amsthm,mathtools,bm}

\newtheorem{remark}{Remark}

\newtheorem*{theorem*}{Theorem}

\def\E{\mathbb{E}}

\newcommand{\SOMMA}[2]{\sum\limits_{#1}^{#2}}

\allowdisplaybreaks

\usepackage[ruled,vlined]{algorithm2e}
\SetKwProg{Fn}{Function}{:}{end}
\SetKw{KwBreak}{break}

\usepackage{graphicx}
\graphicspath{{Images/}}
\usepackage{xcolor}
\usepackage{float}
\usepackage{booktabs}
\usepackage{longtable}
\usepackage{enumitem}
\usepackage{subcaption}
\usepackage{svg}
\svgpath{{figures/}}

\usepackage{tikz}
\usepackage{pgfplots}
\pgfplotsset{compat=1.18}
\usetikzlibrary{
    external,
    positioning,
    fit,
    calc,
    decorations.pathreplacing,
    arrows.meta,
    decorations.text,
    backgrounds,
    shapes.symbols,
    decorations.pathmorphing,
    shapes.geometric,
    shadows
}
\usepackage{cleveref}
\usepackage{csquotes}

\let\oldbibliography\thebibliography
\renewcommand{\thebibliography}[1]{%
  \oldbibliography{#1}%
  \setlength{\itemsep}{0pt}%
  \setlength{\parskip}{0pt}%
  \linespread{0.9}\selectfont%
  \scriptsize%
}

\usepackage{setspace}
\title{Thermodynamics of Forgetting in a Bounded Hopfield Network}

\author[a,b]{Andrea Lepre}
\affiliation[a]{Dipartimento di Matematica, Sapienza Università di Roma, Italy}
\affiliation[b]{GNFM, Istituto Nazionale di Alta Matematica Francesco Severi (INdAM), Roma, Italy}

\abstract{{A memory that keeps learning must also forget. The Hopfield network does not: beyond a
critical number of stored patterns it falls into a state of total confusion and retrieves none of them.
Parisi showed in 1986 that bounding the synaptic strength cures this pathology, since the clipping of the
couplings gradually erases the oldest memories and leaves room for the new ones. In this paper we provide a
thermodynamic description of the resulting network, known as the Hopfield-Parisi model. Starting from the
learning rule, we show that the network is equivalent to a Hopfield model whose patterns are weighted
according to their age, superimposed on a Sherrington-Kirkpatrick spin glass, both governed by the synaptic
bound: a larger bound makes the memory longer, but also noisier. The theory of nonlinear networks developed
in the eighties already provides, through the heuristic replica method, the equations that govern this model at any
temperature, but to the best of our knowledge they have been solved only at zero temperature. We rederive
them at the replica symmetric level through Guerra's interpolation and solve them at finite temperature,
obtaining the phase diagram of the model together with its memory horizon, namely the largest age at which
a pattern can still be retrieved. The solution shows that forgetting acts on the signal alone, the noise
being the same for patterns of every age, and that the horizon is maximal at an intermediate bound which is
essentially insensitive to temperature. Monte Carlo simulations of networks trained with the original
learning rule support these predictions.}}

\begin{document}

\maketitle

\section{Introduction}
\label{sec:intro}
{
Since Hopfield's seminal work \cite{hopfield1982neural}, the statistical mechanics of disordered systems
\cite{mezard1987spin} has proved to be a privileged route toward a quantitative understanding of associative
memories \cite{amit1989modeling,coolen2005theory}. The theory developed by Amit, Gutfreund and Sompolinsky
\cite{amit1985storing,amit1987statistical} turned the Hopfield network into a solvable model, whose phase
diagram splits the plane of the control parameters, the temperature and the load, into regions with
qualitatively different computational capabilities. One feature of that phase diagram is however
unsatisfactory for a memory meant to operate over long times: when the number of stored patterns exceeds a
critical fraction $\alpha_c\simeq0.138$ of the number of neurons, the network does not lose its oldest
memories but all of them at once, and falls into a spin glass state in which no pattern is retrieved. This
collapse is closely related to the catastrophic forgetting suffered by artificial networks trained on a
sequence of tasks \cite{mccloskey1989catastrophic,french1999catastrophic,kirkpatrick2017overcoming}.

A way out was suggested by the observation that biological synapses cannot grow without limit, their
efficacy being confined to a finite range and possibly to a few discrete levels
\cite{petersen1998allornone,oconnor2005graded}. A network that learns continuously with bounded synapses has
to make room for new memories by erasing the old ones, and therefore behaves as a palimpsest, in which only
the most recent traces are legible \cite{nadal1986palimpsest,mezard1986solvable}. In the simplest
realization of this idea, due to Parisi \cite{parisi1986memory}, every coupling receives the Hebbian
increment carried by each new pattern and is clipped whenever its modulus exceeds a bound $A$. 
The simulations of \cite{parisi1986memory} showed that, for a suitable
choice of the bound, the total confusion never occurs and the last $0.04N$ patterns remain retrievable.

The model, which we shall call the Hopfield-Parisi model following \cite{vanhemmen1988forgetful}, was solved
by van Hemmen, Keller and K\"uhn \cite{vanhemmen1988forgetful} within the replica theory of nonlinear neural
networks \cite{vanhemmen1986nonlinear,vanhemmen1987nonlinear,vanhemmen1988nonlinearI,vanhemmen1988nonlinearII}.
In that framework the clipped couplings are expanded on a basis in which every pattern carries a weight that
decays with its age, computed  by regarding the learning stage as a random walk with
reflecting barriers \cite{vanhemmen1987nonlinear}; at zero temperature, the capacity obtained in this way agrees well with the
simulations of \cite{parisi1986memory}. A closely related problem, in which the patterns carry
weights that decrease with their age by construction rather than as an effect of the bound, had been solved
with the replica method in \cite{vanhemmen1987weighted,mezard1986solvable}. 
Learning within bounds has also been studied through signal-to-noise analysis \cite{gordon1987memory},
in a diluted asymmetric version of the model \cite{derrida1987learning} and in further work on forgetful
memories \cite{behn1993multifractality}, and the capacity of bounded synapses exposed to a continuous
stream of memories has become a central question in computational neuroscience
\cite{amit1994learning,fusi2005cascade,fusi2007limits,benna2016computational}.
More recently, the
Hopfield-Parisi model has been revisited by extensive simulations \cite{marinari2019forgetting}, which
provided a finite size estimate of its capacity and showed that the basin of attraction of a pattern shrinks
exponentially with its age.

In spite of this body of work, the thermodynamics of the Hopfield-Parisi model has been explored only in
part. Although its replica symmetric equations hold at any temperature, to the best of our knowledge they
have been solved only at zero temperature, which is also where the simulations of
\cite{parisi1986memory,marinari2019forgetting} were performed, so that the phase diagram in the plane of the
synaptic bound and of the temperature is not known. Moreover, in the spectral approach the noise generated by
the clipping is never written as an energy of its own, but enters the free energy only through the total
squared weight of the nonlinear modes.

In this work we address both points. Our first result is an effective Hamiltonian for the network in its
stationary regime. Expanding the clipped couplings on the characters of the Boolean cube
\cite{odonnell2014analysis} we recover the spectral decomposition of \cite{vanhemmen1987nonlinear}, and we
separate its modes into two groups of different nature: the patterns, each weighted by a profile that
decays over ages of order $A^2$, and a multitude of residual modes, none of which carries a finite weight.
Replacing the latter by a Gaussian coupling, in the spirit of \cite{sompolinsky1986neural}, the
Hopfield-Parisi model becomes the Hopfield network with weighted patterns of \cite{vanhemmen1987weighted},
its weights being now fixed by the learning rule, superimposed on a Sherrington-Kirkpatrick spin glass of
intensity proportional to $A$. 
We test this replacement on directly trained networks, through both the law of the residual couplings and their spectral density. 
The bound thus plays a twofold role: a larger $A$ extends the memory but also increases the noise.

Our second result is a derivation of the thermodynamics of the effective model through Guerra's
interpolation \cite{guerra2002thermodynamic,guerra2003broken}, in the form developed for Hopfield-like
networks \cite{barra2010analogical,agliari2020generalized}. The replica symmetric quenched statistical
pressure we obtain coincides with the one that follows from the replica theory of
\cite{vanhemmen1987nonlinear,vanhemmen1987weighted} once the weights of the Parisi rule are inserted, and it
reduces to the Amit-Gutfreund-Sompolinsky theory for a flat weight profile without glassy term. 
The interpolation recovers this result without analytic continuation in the number of replicas, the replica symmetric assumption being stated explicitly as the self-averaging of the order parameters.
Written in this form, the equations make transparent how the network
forgets: the age of the pattern one tries to retrieve enters only through the weight multiplying the signal,
while the noise is the same for every age, so that old memories are lost because their signal fades, not
because the interference grows.

Our third result is the solution of these equations at finite temperature. We derive the phase diagram in
the plane of the synaptic bound and of the temperature, which displays a retrieval, a spin glass and an
ergodic region, and the memory horizon, namely the largest age at which a pattern can still be retrieved.
The horizon depends non-monotonically on the bound, as a consequence of its twofold role, and is maximal at
$A\simeq0.35$; temperature shortens it by an order of magnitude between $T = 0$ and $T = 0.6$ but leaves the optimal bound almost unchanged. 
At zero temperature we recover the capacity curve of \cite{vanhemmen1988forgetful} everywhere except
close to its end point, where the retrieval region closes at $A_c=0.755$, a slightly larger bound than the
one quoted in \cite{vanhemmen1988forgetful}; as discussed in Section~\ref{sec:numerics}, the difference
concerns the determination of the end point rather than the equations, and is confined to bounds at which
fewer than $10^{-2}N$ patterns are still retrievable.

Finally, we test these predictions against Monte Carlo simulations of networks trained with the original
learning rule, which checks the whole construction at once.

The paper is organized as follows. In Section~\ref{sec:model} we introduce the model and the learning rule;
in Section~\ref{sec:quadratic} we derive the effective Hamiltonian; in Section~\ref{sec:guerra} we obtain its
replica symmetric thermodynamics through Guerra's interpolation; in Section~\ref{sec:numerics} we solve the
resulting equations, present the phase diagrams and compare them with Monte Carlo simulations. Conclusions
and outlooks are drawn in Section~\ref{sec:conclusions}, while technical details are collected in the
Appendices.
}
\section{The model}
\label{sec:model}

 We first recall the Hopfield model and the way it breaks down
when too many patterns are stored.
We consider a network of $N$ binary neurons $\bm\sigma=(\sigma_1,\dots,\sigma_N)\in\{-1,+1\}^N$ that is
exposed to an ordered sequence of $K$ patterns $\bm\xi^1,\dots,\bm\xi^K$, each made of $N$ i.i.d. entries with
\begin{equation}
\label{eq:prob_xi}
\mathbb P(\xi^k_i=+1)=\mathbb P(\xi^k_i=-1)=\dfrac12 .
\end{equation}
The Hopfield model \cite{hopfield1982neural} is defined by the standard quadratic form
\begin{equation}
\mathcal H_N^{\mathrm{HM}}(\bm\sigma|\bm\xi):=-\dfrac12\sum_{\substack{i,j=1\\ i\neq j}}^{N,N}J_{ij}\,\sigma_i\sigma_j,
\qquad
J_{ij}:=\dfrac1N\sum_{k=1}^{K}\xi^k_i\xi^k_j ,
\label{eq:hopfield}
\end{equation}
where the diagonal terms have been removed. Introducing the load of the
network, namely the ratio between the number of patterns and the size of the network in the thermodynamic
limit,
\begin{equation}
\alpha:=\lim_{N\to+\infty}\dfrac{K}{N},
\label{eq:load}
\end{equation}
it is well known that for $\alpha>\alpha_c\simeq0.138$ the network goes into a state of total confusion and
behaves as a spin glass \cite{amit1985storing,amit1989modeling} \footnote{We recall the load only as a term of
comparison. In the model introduced below the network is exposed to an unbounded stream of patterns, so that
\eqref{eq:load} diverges and carries no information.

}.

To prevent the network from overloading and suffering from catastrophic interference, we introduce the concept of {``forgetting''}. In contrast to standard associative memories that preserve all stored information indefinitely until breakdown, the goal here is to dynamically erase older configurations so that synaptic room is continuously made available for newly arriving patterns. Following
\cite{parisi1986memory}, the synaptic couplings are updated each time a new pattern is presented and their
modulus cannot exceed a threshold $A>0$. Denoting by $f_A:\mathbb R\to[-A,A]$ the clipping function
\begin{equation}
f_A(u):=\max\{-A,\min\{u,A\}\},
\label{eq:clipping}
\end{equation}
and starting from $J^{(0)}_{ij}=0$, the couplings after the presentation of the $k$-th pattern are
\begin{equation}
J^{(k)}_{ij}:=f_A\Big(J^{(k-1)}_{ij}+\varepsilon\,\xi^k_i\xi^k_j\Big),
\qquad \varepsilon:=\frac{1}{\sqrt N},
\qquad i\neq j,\quad k=1,\dots,K,
\label{eq:parisi_rule}
\end{equation}
with $J^{(k)}_{ij}=J^{(k)}_{ji}$ and $J^{(k)}_{ii}=0$. We write $J_{ij}:=J^{(K)}_{ij}$ for the couplings at
the end of the learning stage. The prescription is as local as the Hebbian one, since the update of a synapse
uses only the activity of the two neurons it connects, and it differs from it in a single respect: the
information about a pattern is not stored once and for all, but is gradually pushed out of the synapse by the
patterns that follow, until the clipping has erased it.

\begin{remark}
\label{rem:scaling}
The increment in \eqref{eq:parisi_rule} is of order $N^{-1/2}$, whereas the Hebbian couplings
\eqref{eq:hopfield} are normalized by $N^{-1}$. This is the scaling of \cite{parisi1986memory}, and it is the
one that keeps the bound $A$ of order one: with this choice the coupling explores the whole interval $[-A,A]$,
the clipping is active on a finite fraction of the updates, and the number of patterns that leave a trace in
the synapses is extensive. The two conventions therefore differ by a global factor, since for $A\to+\infty$
the clipping never intervenes and \eqref{eq:parisi_rule} returns the Hebbian prescription
\eqref{eq:hopfield} multiplied by $\sqrt N$. The learning rule fixes the couplings, so that
factor has to be reabsorbed when the energy is defined, which is done in Section \ref{sec:quadratic}.
\end{remark}

Since the network is supposed to retrieve recent inputs, it is convenient to label the patterns by their age
rather than by the order of presentation. After $K$ presentations we set
\begin{equation}
\bm\xi^{\mu}:=\bm\xi^{K-\mu},\qquad x:=\dfrac{\mu}{N},
\label{eq:age}
\end{equation}
so that $\mu=0$ is the most recent pattern and $x$ is the rescaled age, which becomes a continuous variable in
the thermodynamic limit.

{Throughout the paper we work in the stationary regime, in which the network has been learning long
enough for the initial condition $J^{(0)}_{ij}=0$ to be forgotten. In this regime the law of the couplings no
longer depends on the number $K$ of presentations; the condition under which the regime is reached, $K\gg A^2N$, is derived in
Section~\ref{sec:hamiltonian}.}

\section{From the learning rule to the quadratic form}
\label{sec:quadratic}

The rule \eqref{eq:parisi_rule} defines the couplings implicitly, as a composition of clippings. Following \cite{vanhemmen1987nonlinear,vanhemmen1988forgetful}, we expand them on the basis that diagonalizes them, compute the weights of the resulting modes, and replace the residual ones by a Gaussian coupling, which leads to the effective Hamiltonian \eqref{eq:H_final}.

\subsection{The spectral decomposition of the couplings}
\label{sec:spectral}

Iterating \eqref{eq:parisi_rule}, the coupling at the end of the learning stage is a fixed function of the
signs $s^\mu_{ij}:=\xi^\mu_i\xi^\mu_j$ alone, the patterns being {labeled} by their
age as in \eqref{eq:age}, so the composition below starts from the
oldest,
\begin{equation}
J_{ij}=F(\bm s_{ij}),\qquad
F(\bm s)=f_A\Big(\cdots f_A\big(f_A(\varepsilon s_{K-1})+\varepsilon s_{K-2}\big)\cdots+\varepsilon s_0\Big),
\label{eq:F}
\end{equation}
and $F$ is the same for every pair. Writing $\bm x\circ\bm y$ for the componentwise product of two vectors and
$\bm\xi_i:=(\xi^0_i,\dots,\xi^{K-1}_i)$ for the information available at site $i$, one has
$\bm s_{ij}=\bm\xi_i\circ\bm\xi_j$: the coupling depends on the two sites only through that product, which is
the locality and invariance condition of \cite{vanhemmen1987nonlinear}. Two properties of $F$ will be used.
\begin{itemize}
\item [(i)]The first is that $F$ is odd, $F(-\bm s)=-F(\bm s)$, which follows by induction from \eqref{eq:F} since $f_A$
is odd and $J^{(0)}_{ij}=0$. 
\item [(ii)] The second is that, for a fixed pair, the components of $\bm s_{ij}$ are i.i.d.
and uniform on $\{-1,1\}$, because $\mathbb P(\xi^\mu_i\xi^\mu_j=1)=\mathbb P(\xi^\mu_i=\xi^\mu_j)=1/2$ and the pairs
$(\xi^\mu_i,\xi^\mu_j)$ are independent for different $\mu$.\footnote{The vectors $\bm s_{ij}$ of different pairs
are not jointly independent, since $s^\mu_{ij}s^\mu_{jl}s^\mu_{li}=1$ for any triple of sites. This does not affect
the expansion below, which is an identity for each pair separately, and it will be taken into account when
computing moments.}
\end{itemize}

Now, any function on the discrete cube $\{-1,1\}^K$ is a linear combination of the products of its arguments, a
fact known as the Fourier or Walsh expansion \cite{odonnell2014analysis}. For every subset
$\rho\subseteq\{0,\dots,K-1\}$ we define 
\begin{equation}\label{character}
\chi_\rho(\bm s):=\prod_{\mu\in\rho}s_\mu, 
\end{equation}
with $\chi_\emptyset\equiv1$.
These functions are called the \emph{characters} of the cube because each of them turns the product of two
configurations into the product of the two values, $\chi_\rho(\bm s\circ\bm s')=\chi_\rho(\bm s)\chi_\rho(\bm
s')$; equivalently they satisfy $\chi_\rho\chi_{\rho'}=\chi_{\rho\triangle\rho'}$ and
$\E_{\bm s}\chi_\rho=\delta_{\rho\emptyset}$, where $\E_{\bm s}$ denotes the average over $\bm s$ uniform on
$\{-1,1\}^K$. The two identities say that the characters are orthonormal, so that expanding on them is just a
change of orthonormal basis in a finite dimensional space and the coefficients are the corresponding scalar
products,
\begin{equation}
F(\bm s)=\sum_{\rho\subseteq\{0,\dots,K-1\}}\hat F(\rho)\,\chi_\rho(\bm s),
\qquad
\hat F(\rho)=\E_{\bm s}\big[F(\bm s)\chi_\rho(\bm s)\big].
\label{eq:walsh}
\end{equation}
Two consequences follow at once. Since $\chi_\rho(-\bm s)=(-1)^{|\rho|}\chi_\rho(\bm s)$, the parity
of $F$ makes the coefficients with $|\rho|$ even vanish, and orthonormality gives Parseval's identity,
\begin{equation}
\sum_\rho\hat F(\rho)^2=\E_{\bm s}\big[F(\bm s)^2\big].
\label{eq:parseval}
\end{equation}

The reason why \eqref{eq:walsh} is the useful expansion is that the characters factorize over the two sites.
Setting 
\begin{equation}v_\rho(\bm\xi_i):=\prod_{\mu\in\rho}\xi^\mu_i,
\end{equation}
which is the same character evaluated at
the history of a single site and is again a $\pm1$ variable, one has
$\chi_\rho(\bm\xi_i\circ\bm\xi_j)=v_\rho(\bm\xi_i)v_\rho(\bm\xi_j)$, and therefore
\begin{equation}
J_{ij}=\sum_{|\rho|\ \mathrm{odd}}\hat F(\rho)\,v_\rho(\bm\xi_i)\,v_\rho(\bm\xi_j).
\label{eq:J_exact}
\end{equation}
This is the spectral decomposition of \cite{vanhemmen1987nonlinear}; see also
\cite{vanhemmen1988nonlinearI}. Read as a function of the two site informations, the coupling has been
diagonalized: the vectors $\bm v_\rho=(v_\rho(\bm\xi_1),\dots,v_\rho(\bm\xi_N))$ play the role of
eigenvectors and the numbers $\hat F(\rho)$ that of eigenvalues. All the randomness sits in the $\bm v_\rho$,
while the coefficients are deterministic. 

For $\rho=\{\mu\}$, $\mu \in \{0, \dots, K-1\}$, the vector $\bm v_\rho$ is the pattern
$\bm\xi^\mu$ itself, and for $|\rho|\ge3$ it is the site by site product of an odd number of patterns. From here on the two families of indices are distinguished by the symbol alone: $\rho$ denotes
a general subset, $\mu$ a single pattern.

\subsection{The Hamiltonian}
\label{sec:hamiltonian}

The decomposition \eqref{eq:J_exact} is exact but it is not yet an energy, because the scale is still to be
fixed. As anticipated in Remark \ref{rem:scaling}, 
the Hamiltonian has to be of order $N$, so that the free energy per neuron has a finite limit.

The size of the coefficients tells us which power of $N$ to divide by, and for the most recent pattern it can
be read off directly, {once the law of a single coupling in the stationary regime is known. To find
it, fix a pair $(i,j)$ and assume for simplicity that $A\sqrt N=L$ is a positive integer, so that along the
learning stage $J^{(k)}_{ij}$ takes values on the lattice $\{-L\varepsilon,\dots,L\varepsilon\}$ of $2L+1$
points. Since the signs $\xi^k_i\xi^k_j$ are independent and uniform for different $k$, the sequence
$J^{(k)}_{ij}$ is a Markov chain: at each step the coupling moves up or down by $\varepsilon$ with probability
$1/2$, except at the two endpoints $\pm A$, where a step pushing outwards is canceled by the clipping and the
coupling stays put. This is a symmetric random walk with reflecting barriers \cite{feller1968introduction}.
Its transition matrix is symmetric, hence doubly stochastic, so the uniform distribution on the $2L+1$ sites
is invariant; the chain is irreducible, and it is aperiodic because of the holding probability at the
endpoints, so its law converges to the uniform distribution as $k\to\infty$. The relaxation time is set by the
spectral gap of the walk and is of order $A^2N$, which is the condition $K\gg A^2N$ for the stationary regime
anticipated in Section~\ref{sec:model}. In this regime the first two moments of a coupling are
\begin{equation}
\E\big[J_{ij}\big]=0,
\qquad
\E\big[J_{ij}^2\big]=\frac1{2L+1}\sum_{l=-L}^{L}(l\varepsilon)^2=\frac{A(A+\varepsilon)}{3}
\;\longrightarrow\;\frac{A^2}{3},
\label{eq:stationary}
\end{equation}
and this is the regime considered in \cite{vanhemmen1988forgetful}, where the same Markov chain is used to
compute the weights of the patterns. In particular,} $J^{(K-1)}_{ij}$ is uniform on {this}
lattice and independent of $s^0_{ij}$, so that, denoting by $U$ that uniform variable,
\begin{equation}
\hat F(\{0\})=\E_{\bm s}\big[F\,s_0\big]
=\frac12\Big(\E\big[f_A(U+\varepsilon)\big]-\E\big[f_A(U-\varepsilon)\big]\Big)
=\frac{2{L}}{2{L}+1}\,\varepsilon .
\label{eq:Fhat_last}
\end{equation}
{Indeed, $f_A(U+\varepsilon)=U+\varepsilon$ on all the $2L+1$ sites except the topmost one, where the
clipping gives back $U=A$, so that $\E[f_A(U+\varepsilon)]=\E[U]+\frac{2L}{2L+1}\varepsilon$, and symmetrically
$\E[f_A(U-\varepsilon)]=\E[U]-\frac{2L}{2L+1}\varepsilon$.} The coefficient of a single pattern is therefore of order
$\varepsilon=N^{-1/2}$, and multiplying it by $\sqrt N$ gives a quantity of order one, equal to $1$ in the
limit. Setting accordingly
\begin{equation}
\mathcal H_N^{\mathrm{HPM}}(\bm\sigma|\bm\xi):=-\frac1{2\sqrt N}\sum_{i\neq j}J_{ij}\sigma_i\sigma_j,
\qquad
\Lambda_\rho:=\sqrt N\,\hat F(\rho),
\label{eq:H_def}
\end{equation}
and using $\sum_{i\neq j}a_ia_j=(\sum_ia_i)^2-\sum_ia_i^2$ with $a_i=v_\rho(\bm\xi_i)\sigma_i$ and $a_i^2=1$,
we obtain the quadratic form
\begin{equation}
\mathcal H_N^{\mathrm{HPM}}(\bm\sigma|\bm\xi)=-\frac1{2N}\sum_{|\rho|\ \mathrm{odd}}\Lambda_\rho
\Big(\sum_{i=1}^Nv_\rho(\bm\xi_i)\,\sigma_i\Big)^2+\mathrm{const},
\label{eq:H_exact}
\end{equation}
which is eq.~(10) of \cite{vanhemmen1987nonlinear}. The constant equals
$\frac12\sum_\rho\Lambda_\rho=\frac12A\sqrt N$, because evaluating \eqref{eq:walsh} at $\bm s=(1,\dots,1)$
gives $\sum_\rho\hat F(\rho)=F(1,\dots,1)=\min\{K\varepsilon,A\}=A$ as soon as $K\ge A\sqrt N$; it is
therefore subextensive and plays no role in the limit. The model has thus become a Hopfield model whose patterns are the
vectors $\bm v_\rho$, each entering with its own weight $\Lambda_\rho$, and it remains to say how large those
weights are.

\subsection{The weights}
\label{sec:weights}

The weight of a pattern of age $\mu$ is the one carried by the singleton $\{\mu\}$, which we abbreviate
as $\Lambda_\mu:=\Lambda_{\{\mu\}}$, so that the weights indexed by $\mu$ are the restriction of the family
$\Lambda_\rho$ to the singletons{. It equals} $\Lambda_{\mu}=\sqrt N\,\E_{\bm s}[F\,s_\mu]$, that is the
residual correlation between the coupling and the pattern that was written $\mu$ steps earlier. In the
stationary regime it depends on the age only through $x=\mu/N$, and for $\mu=\lfloor xN\rfloor$ it converges
to
\begin{equation}
\Lambda(x)=\frac{8}{\pi^2}\sum_{n\ \mathrm{odd}}\frac1{n^2}
\exp\Big(-\frac{n^2\pi^2}{8A^2}\,x\Big),
\label{eq:Lambda}
\end{equation}
which is eq.~(6) of \cite{vanhemmen1988forgetful}. 
Consistently with \eqref{eq:Fhat_last} one has $\Lambda(0)=1$, since the
coefficients $8/(\pi^2n^2)$ sum to one.

\begin{figure}[t]
\centering
\includegraphics[width=\linewidth]{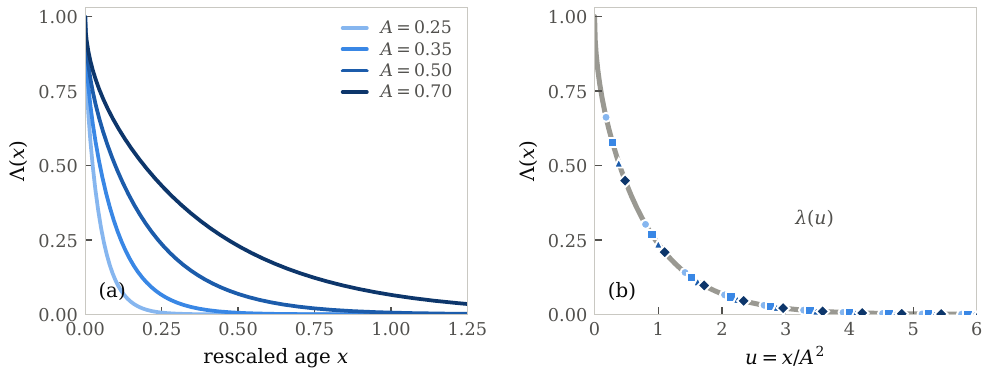}
\caption{The weight profile \eqref{eq:Lambda}. \emph{(a)} $\Lambda$ against the rescaled age $x$, for four
values of the synaptic bound $A$. The most recent pattern always carries weight one, and the trace a pattern
leaves in the synapses is erased on the age scale $x\sim A^2$. \emph{(b)} The same curves against
$u=x/A^2$: the bound enters \eqref{eq:Lambda} only through this combination, so that all the profiles collapse
on a single function $\lambda(u)$ (grey line), and {$A^2$} acts as the unit in which the age of a pattern is
measured.}
\label{fig:lambda}
\end{figure}

The origin of \eqref{eq:Lambda} is easy to describe. As explained in {Section~\ref{sec:hamiltonian}}, each
coupling performs a random walk with steps $\pm\varepsilon$ between two reflecting barriers at $\pm A$, and
after $\mu$ steps the accumulated variance is $\mu\varepsilon^2=x$, which is why $x$ is the natural time. The
trace left by a pattern is the linear response of the coupling to an initial shift of size $\varepsilon$, and
such a response relaxes as the heat kernel on an interval with reflecting ends, that is with the modes
$\exp(-n^2\pi^2x/(8A^2))$ and the amplitudes $8/(\pi^2n^2)$. The kernel decays exponentially on the age scale
$x\sim A^2$, and this is the forgetting mechanism. Figure \ref{fig:lambda} shows the profile of $\Lambda(x)$ for a few values
of the bound.

The remaining weights are not needed one by one, only their total. Since $J_{ij}=F(\bm s_{ij})$ with
$\bm s_{ij}$ uniform, we have $\E_{\bm s}[F^2]=\E[J_{ij}^2]\to A^2/3$ by \eqref{eq:stationary}, and
\eqref{eq:parseval} gives
\begin{equation}
\frac1N\sum_{|\rho|\ge3}\Lambda_\rho^2
=\E_{\bm s}\big[F^2\big]-\frac1N\sum_{\mu}\Lambda_{\mu}^2
\ \longrightarrow\
\sigma_\eta^2:=\frac{A^2}{3}-\int_0^\infty\Lambda(x)^2dx\simeq0.052\,A^2 .
\label{eq:sigma}
\end{equation}
Moreover $\Lambda_\rho\to0$ as $N\to+\infty$ for every $\rho$ with $|\rho|\ge3$, as stated in
\cite{vanhemmen1988forgetful}. The residual part of the coupling is thus made of very many modes, each with a
vanishing weight, whose squared weights add up to the finite quantity $\sigma_\eta^2$.

\subsection{The effective model}
\label{sec:effective}

The decomposition \eqref{eq:H_exact} therefore contains two groups of modes that play very different roles.
Those with $|\rho|=1$ are the patterns themselves, each carrying the macroscopic weight \eqref{eq:Lambda};
those with $|\rho|\ge3$ are the residual part just described, of which only the total \eqref{eq:sigma}
matters. It is natural to keep the square for the first group, where a single mode can dominate, and to reopen
it for the second, where none does.

We define accordingly
\begin{equation}
\eta_{ij}:=\frac1{\sqrt N}\sum_{|\rho|\ge3}\Lambda_\rho\,v_\rho(\bm\xi_i)v_\rho(\bm\xi_j),
\label{eq:eta}
\end{equation}
so that $\sum_{|\rho|\ge3}\Lambda_\rho v_\rho(\bm\xi_i)v_\rho(\bm\xi_j)=\sqrt N\,\eta_{ij}$. Using once more
$\big(\sum_ia_i\big)^2=\sum_{i\neq j}a_ia_j+N$ with $a_i=v_\rho(\bm\xi_i)\sigma_i$, the second group of modes
becomes a pair interaction,
\begin{equation}
-\frac1{2N}\sum_{|\rho|\ge3}\Lambda_\rho\Big(\sum_{i=1}^Nv_\rho(\bm\xi_i)\,\sigma_i\Big)^2
=-\frac1{\sqrt N}\sum_{i<j}\eta_{ij}\,\sigma_i\sigma_j-\frac12\sum_{|\rho|\ge3}\Lambda_\rho .
\label{eq:reopen}
\end{equation}
Added to the constant of \eqref{eq:H_exact}, the last term leaves $\frac12\sum_\mu\Lambda_{\mu}$, which is
of order $N$ because $N^{-1}\sum_\mu{\Lambda_\mu}\to\int_0^\infty\Lambda(x)dx=\frac23A^2$.\footnote{At variance
with the constant of \eqref{eq:H_exact} this one is extensive and shifts the free energy per neuron by a known
amount; being independent of $\bm\sigma$ it does not affect the Gibbs measure, and we drop it, to be restored
if the energy itself rather than the phase diagram is of interest.}
\begin{figure}[t]
\centering
\includegraphics[width=\linewidth]{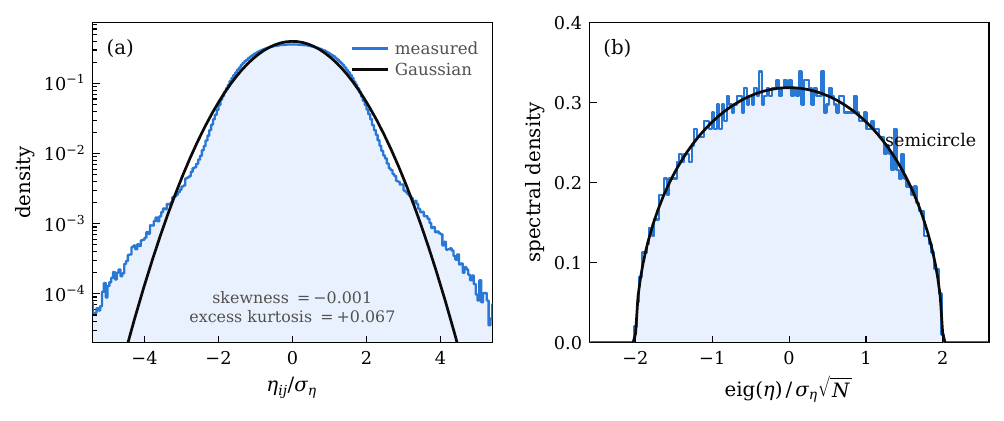}
\caption{Test of the Gaussian replacement of the residual couplings on a
directly simulated network. A blank network of $N=3000$ neurons is trained
with the rule \eqref{eq:parisi_rule} at $A=0.73$, so that a synapse has
${L}=A\sqrt N=40$ lattice sites available, and $1.3\times10^4$ patterns are
presented, that is up to a rescaled age $x=8A^2$. Since the slowest mode of
the reflecting walk decays as $\exp(-\pi^2x/8A^2)$, this leaves a relaxation
factor of order {$5\cdot10^{-5}$} and the network is in the stationary regime of
{Section~\ref{sec:hamiltonian}}. The residual coupling is then obtained as
$\eta_{ij}=J_{ij}-N^{-1/2}\sum_\mu\Lambda_\mu\xi^\mu_i\xi^\mu_j$, the weights
$\Lambda_\mu$ being measured on the same network as
$\sqrt N\,\mathbb E[J_{ij}\xi^\mu_i\xi^\mu_j]$, so that $\eta$ is exactly the
part of the coupling carried by the modes with $|\rho|\ge3$. Its variance is
$0.0542\,A^2$, against the asymptotic value $0.0522\,A^2$ of
\eqref{eq:sigma}; the residual $4\%$ is the $O(1/{L})$ lattice correction and
vanishes as ${L}\to\infty$.
\emph{(a)} Law of $\eta_{ij}/\sigma_\eta$ against a standard Gaussian, on a
logarithmic vertical scale so that the tails are displayed rather than
implied. The bulk is Gaussian to the eye, with skewness $-0.001$ and excess
kurtosis $+0.067$; beyond about $2.5\,\sigma_\eta$ the measured tails lie
slightly above the Gaussian ones.
\emph{(b)} Spectral density of $\eta/(\sigma_\eta\sqrt N)$ against the
semicircle law on $[-2,2]$, the measured edge falling at $1.986$. This is the
stronger of the two tests: the marginal law in \emph{(a)} constrains a single
entry, whereas the spectral density is sensitive to the correlations among
entries, which are what the replacement assumes away. 
}
\label{fig:gaussian}
\end{figure}
The couplings \eqref{eq:eta} are {centered}, and the orthonormality of the characters gives their moments,
\begin{equation}
\E\,\eta_{ij}=0,
\qquad
\E\,\eta_{ij}^2=\frac1N\sum_{|\rho|\ge3}\Lambda_\rho^2\longrightarrow\sigma_\eta^2,
\qquad
\E\big[\eta_{ij}\eta_{lr}\big]=0,
\qquad
\E\big[\eta_{ij}\,\xi^\mu_i\xi^\mu_j\big]=0,
\label{eq:eta_moments}
\end{equation}
the third holding for $\{i,j\}\neq\{l,r\}$ and the fourth for every $\mu$. The last two identities say that
distinct pairs are uncorrelated and that no residual correlation is left with the patterns kept in the first
group. This, together with the fact that no single mode contributes to $\eta_{ij}$ with a finite weight, is
what suggests treating the residual coupling as Gaussian: we replace $\eta_{ij}$ by $\sigma_\eta z_{ij}$, with
$z_{ij}$ i.i.d. standard Gaussian variables, a substitution that reproduces \eqref{eq:eta_moments} exactly, see Fig. \ref{fig:gaussian}.

One last step is a matter of convenience rather than of approximation. The sum over the first group runs over
all the patterns ever presented, and by {Section~\ref{sec:hamiltonian}} the stationary regime requires
$K\gg A^2N$, so that the rescaled age extends well beyond the scale $A^2$ on which the weights decay. In order
to work at finite $N$ with a fixed number of patterns per neuron we introduce a cutoff $\bar x$ on the age,
keeping the $P:=\lfloor\bar xN\rfloor$ patterns younger than it, and we let $\bar x\to+\infty$ at the end. What
is discarded is a total squared weight $\int_{\bar x}^\infty\Lambda(x)^2dx$, exponentially small in
$\bar x/A^2$: the coefficients $8/(\pi^2n^2)$ in \eqref{eq:Lambda} sum to one and each exponential is bounded
by the first, whence $\Lambda(x)\le e^{-\pi^2x/(8A^2)}$ and
\begin{equation}
\int_{\bar x}^{\infty}\Lambda(x)^2dx\ \le\ \frac{4A^2}{\pi^2}\,e^{-\pi^2\bar x/(4A^2)} .
\label{eq:tail}
\end{equation}

Collecting the three steps, we arrive at the effective Hamiltonian of the Hopfield-Parisi model,
\begin{equation}
\mathcal H_{N,\bar x}^{\rm eff}(\bm\sigma|\bm\xi,\bm z)=
-\frac1{2N}\sum_{\mu=0}^{P-1}\Lambda_\mu\Big(\sum_{i=1}^N\xi^\mu_i\sigma_i\Big)^2
-\frac{\sigma_\eta}{\sqrt N}\sum_{i<j}z_{ij}\sigma_i\sigma_j,
\qquad \Lambda_\mu=\Lambda(\mu/N),
\label{eq:H_final}
\end{equation}
where the patterns are {labeled} by their age as in \eqref{eq:age}. The model is a Hopfield network whose
patterns are weighted by their age, superimposed on a Sherrington-Kirkpatrick model of intensity
$\sigma_\eta$: the first term is what the clipping preserves of the memory, the second is the noise the
clipping produces, and both are governed by the single parameter $A$. 

\begin{remark}\label{rem:gaussian}
The quadratic form \eqref{eq:H_exact} is exact, and the value \eqref{eq:Lambda} of the weights is quoted from
\cite{vanhemmen1988forgetful}. The only assumption of the construction is the replacement of $\eta_{ij}$ by a
Gaussian coupling.
{In Remark~\ref{rem:gaussianRS} we show that, at the replica symmetric level, the two treatments
give the same pressure.} 
\end{remark}

\section{Analytical Results}
\label{sec:guerra}

The aim of this section is to compute the quenched statistical pressure of the effective model
\eqref{eq:H_final}.
The route we follow is Guerra's interpolation
\cite{guerra2003broken,guerra2002thermodynamic}, in the form in which it has been adapted to Hopfield-like
networks {\cite{barra2010analogical,agliari2020generalized}}.  
Before setting up this machinery, however, we have to say precisely which parameters the model
depends on, which macroscopic observables describe its states, and which thermodynamic functionals we are
going to compute.

\subsection{Control and Order parameters}
\label{sec:control}
The model is ruled by the following \emph{control
parameters}, all held fixed in the thermodynamic limit:

\begin{itemize}
\item the \emph{synaptic bound} $A>0$ of the learning rule \eqref{eq:parisi_rule}. It enters the effective model \eqref{eq:H_final} twice: through the weight profile
$\Lambda(x)$ of \eqref{eq:Lambda}, and through the intensity
$\sigma_\eta$ of the glassy term.

\item the \emph{inverse temperature} $\beta>0$, tuning the stochasticity of the neurons in the usual way:
   for $\beta\to0$ the neurons are assigned random states, while for $\beta\to\infty$ the dynamics drives them
   deterministically toward the minima of the energy.
We shall also write $T:=\beta^{-1}$ when reporting phase diagrams.

\item the \emph{age of the target} $x_0$. In the Hopfield model all the patterns are equivalent and one may single out any of them as the candidate for
retrieval. Here they are not, each of them carrying the weight of its own age, so that the question is not
only whether the network retrieves, but how far back in time it can do so. We therefore fix an age
$x_0\in[0,\bar x)$, we set $\mu_0:=\lfloor x_0N\rfloor$, and we take $\bm\xi^{\mu_0}$ as the target pattern.
\end{itemize}

A comment is in order. The model lives by construction in the high storage regime, and in fact
beyond it, since the number of patterns per neuron diverges with $\bar x$.
 What keeps the thermodynamics well
defined is not their number but their weights, which are both integrable and square integrable,
\begin{equation}
\int_0^{+\infty}\Lambda(x)\,dx=\frac{2}{3}A^2,
\qquad
\int_0^{+\infty}\Lambda(x)^2dx\simeq0.281\,A^2 .
\label{eq:weighted_loads}
\end{equation}
Now, the macroscopic observables we need are of three kinds.

The retrieval order parameter is the \emph{Mattis magnetization}, that is the normalized overlap between a
pattern and the configuration of the neurons,
\begin{equation}
m_\mu(\bm\sigma|\bm\xi^\mu):=\frac1N\SOMMA{i=1}{N}\xi^\mu_i\sigma_i ,
\qquad \mu=0,\dots,P-1 ,
\label{eq:mattis}
\end{equation}
and we write $m:=m_{\mu_0}$ for the one carried by the target. 

Since the disorder is quenched and \eqref{eq:H_final} contains a Sherrington-Kirkpatrick component, the glassy
character of the equilibrium measure has to be monitored too, and this is the role of the two replica overlap
of the neurons, also known as the \emph{Edwards-Anderson overlap},
\begin{equation}
q_{ab}(\bm\sigma):=\frac1N\SOMMA{i=1}{N}\sigma^{(a)}_i\sigma^{(b)}_i ,
\label{eq:qab}
\end{equation}
where $a$ and $b$ label two replicas of the system, namely two configurations drawn independently from the
same realization of the disorder; notice that $q_{aa}\equiv1$.

The third family of order parameters is produced by the non-condensed patterns, and it is here that the
Hopfield-Parisi model departs from the standard one. {As we shall see} in Section \ref{sec:signalnoise}, the quadratic
terms of \eqref{eq:H_final} with $\mu\neq\mu_0$ are linearized by a Hubbard-Stratonovich transformation, each
of them contributing one analog variable $\tau_\mu$ coupled to the neurons with strength
$\sqrt{\beta\Lambda_\mu/N}$, and the natural order parameter is the two replica overlap of these variables,
\begin{equation}
p_{ab}:=\frac1N\sum_{\mu\neq\mu_0}\Lambda_\mu\,\tau^{(a)}_\mu\tau^{(b)}_\mu .
\label{eq:pab}
\end{equation}
Two features of \eqref{eq:pab} deserve a comment. The first is the weight $\Lambda_\mu$: 
each of the non-condensed patterns contributes in proportion to what the synapses still remember of it, so that the weight measures how many patterns are effectively interfering. The second is the normalization. One would
normally divide by the number of non-condensed patterns, but that is not possible here: they are
$P-1=\lfloor\bar xN\rfloor-1$ with $\bar x\to+\infty$, while the weights are summable by
\eqref{eq:weighted_loads}, so such a normalization would send \eqref{eq:pab} to zero. Dividing by $N$ is the
choice that keeps the overlap finite.

\subsection{Partition function and quenched statistical pressure}
\label{sec:thermo}

In terms of \eqref{eq:mattis} the effective Hamiltonian \eqref{eq:H_final} reads
\begin{equation}
-\beta\,\mathcal H_{N,\bar x}^{\rm eff}(\bm\sigma|\bm\xi,\bm z)
=\frac{\beta N}{2}\SOMMA{\mu=0}{P-1}\Lambda_\mu\,m_\mu^2
+\frac{\beta\,\sigma_\eta}{\sqrt N}\sum_{i<j}z_{ij}\,\sigma_i\sigma_j ,
\label{eq:H_compact}
\end{equation}
and the statistical mechanics of the model is built in the standard way. The partition function at finite size
$N$ and fixed $\bar x$ is
\begin{equation}
\mathcal Z_{N, \bar x, \beta}(\bm\xi,\bm z):=\sum_{\{\bm\sigma\}}
\exp\big[-\beta\,{\mathcal H_{N,\bar x}^{\rm eff}}(\bm\sigma|\bm\xi,\bm z)\big],
\label{eq:Z}
\end{equation}
where $\sum_{\{\bm\sigma\}}$ runs over the $2^N$ configurations of the neurons and
$\exp[-\beta\mathcal H^{\rm eff}]$ is the Boltzmann factor\footnote{Both the patterns and the Gaussian couplings are
quenched, that is frozen on the time scale over which the neurons equilibrate, so that the average over them
has to be taken after the logarithm.}. Writing $\E:=\E_{\bm\xi}\E_{\bm z}$ for this quenched expectation, the
\emph{quenched statistical pressure} at finite size is
\begin{equation}
\mathcal A_{N, \bar x}(\beta,A,x_0):=\frac1N\,\E\big[\ln{\mathcal Z_{N,\bar x,\beta}}(\bm\xi,\bm z)\big].
\label{eq:pressure_N}
\end{equation}
The object we compute is {the limit
\begin{equation}
\mathcal A(\beta,A,x_0):=\lim_{\bar x \to \infty}\lim_{N\to+\infty}\mathcal A_{N,\bar x}(\beta,A,x_0),
\label{eq:pressure}
\end{equation}
in which the thermodynamic limit, assumed to exist and to be unique,}\footnote{Existence and
uniqueness of the quenched statistical pressure are classical for the Sherrington-Kirkpatrick model
\cite{guerra2002thermodynamic,TALAGRAND200169}; for Hopfield-like models the corresponding statements are
more recent \cite{agliari2026infinite}, and we assume them here.} {is taken first, at fixed
cutoff, and the cutoff is removed afterwards, as anticipated in Section~\ref{sec:effective};} the free energy
per neuron is $-\mathcal A/\beta$. The Boltzmann-Gibbs measure on the configurations is
\begin{equation}
\mathbb P_\beta(\bm\sigma|\bm\xi,\bm z)
:=\frac{e^{-\beta{\mathcal H_{N,\bar x}^{\rm eff}}(\bm\sigma|\bm\xi,\bm z)}}{{\mathcal Z_{N,\bar x,\beta}}(\bm\xi,\bm z)},
\label{eq:gibbs}
\end{equation}
and for an observable $\mathcal O(\bm\sigma)$ we denote by
\begin{equation}
\omega(\mathcal O):=\sum_{\{\bm\sigma\}}\mathbb P_\beta(\bm\sigma|\bm\xi,\bm z)\,\mathcal O(\bm\sigma),
\qquad
\langle\mathcal O\rangle:=\E\big[\omega(\mathcal O)\big],
\label{eq:omega}
\end{equation}
the Boltzmann average at fixed disorder and its quenched counterpart. When several replicas are involved,
$\omega$ stands for the product measure over them. 
It is convenient to abbreviate the
values the order parameters take in the thermodynamic limit as
\begin{equation}
\bar m:=\langle m\rangle,\qquad \bar q:=\langle q_{12}\rangle,\qquad \bar p:=\langle p_{12}\rangle .
\label{eq:RSvalues}
\end{equation}

The interpolation yields identities that hold at finite $N$, but the expressions it produces close only if the order parameters
do not fluctuate in the thermodynamic limit. This is the content of the replica symmetric assumption, which we
state in the form we shall use: for every order parameter $X\in\{m,\ q_{12},\ p_{12}\}$, denoting by
$\mathbb P_N(X)$ the distribution induced on it by \eqref{eq:gibbs} and by the disorder,
\begin{equation}
\lim_{N\to+\infty}\mathbb P_N(X)=\delta\big(X-\langle X\rangle\big),
\label{eq:RS}
\end{equation}
namely each order parameter self-averages on a deterministic value. 

\begin{remark}
\label{rem:RSstatus}
It is worth being explicit about the status of \eqref{eq:RS}, since it would be easy to overstate the rigor
of what follows. Guerra's interpolation avoids the analytic continuation in the number of replicas required
by the replica trick, but the computation can be closed only under \eqref{eq:RS}, which is assumed rather
than proved and has to hold not only for the Boltzmann-Gibbs measure but along the whole interpolation
introduced in Section~\ref{sec:interpolating}. The results of this section are therefore replica symmetric.
The replica symmetric solution of Hopfield-like models is moreover expected to become unstable at low
temperature and large load, below the de Almeida-Thouless line \cite{de1978stability,albanese2024almeida},
so that the low temperature portions of the boundaries reported below are to be read as approximate; we
return to this point in the Conclusions.
\end{remark}

\subsection{Signal, noise and the Gaussian representation}
\label{sec:signalnoise}

The first step is the classical separation between the pattern we want to retrieve and all the others.
Writing ${\Lambda_\star}:=\Lambda_{\mu_0}=\Lambda(x_0)$ for the weight of the target and splitting the sum in
\eqref{eq:H_compact} accordingly,
\begin{equation}
-\beta\,\mathcal H_{N, \bar x}^{\rm eff}(\bm\sigma|\bm\xi,\bm z)
=\underbrace{\frac{\beta N}{2}\,{\Lambda_\star}\,m^2}_{\textit{signal}}
\;+\;\underbrace{\frac{\beta N}{2}\sum_{\mu\neq\mu_0}\Lambda_\mu\,m_\mu^2
\;+\;\frac{\beta\sigma_\eta}{\sqrt N}\sum_{i<j}z_{ij}\,\sigma_i\sigma_j}_{\textit{noise}} .
\label{eq:split}
\end{equation}
The signal is a single macroscopic term, and its weight ${\Lambda_\star}$ is the only place where the age of the
target enters. The noise
has two components of different nature, the $P-1$ non-condensed patterns, each entering through a square and
which we assume, {as is customary}, to be Gaussian   \footnote{The replacement is supported by the Carmona-Hu universality theorem for spin glass models \cite{carmona2006universality}, in the version extended to bipartite mean-field spin glasses
\cite{genovese2012universality}.}, and the glassy coupling, which is already a pair
interaction.

Each non-condensed square is linearized by the Hubbard-Stratonovich identity
\begin{equation}
\exp\Big[\frac12\,a\,y^2\Big]=\int\mathcal D\tau\,\exp\big[\sqrt a\,\tau\,y\big],
\qquad
\mathcal D\tau:=\frac{d\tau}{\sqrt{2\pi}}\,e^{-\tau^2/2},
\label{eq:HS}
\end{equation}
applied with $a=\beta N\Lambda_\mu$ and $y=m_\mu$, which introduces one analog variable per non-condensed
pattern,
\begin{equation}
\exp\Big[\frac{\beta N}{2}\Lambda_\mu\,m_\mu^2\Big]
=\int\mathcal D\tau_\mu\,
\exp\Big[\sqrt{\frac{\beta\Lambda_\mu}{N}}\,\tau_\mu\SOMMA{i=1}{N}\xi^\mu_i\sigma_i\Big] ,
\label{eq:HS_applied}
\end{equation}
these being the variables entering \eqref{eq:pab}. The partition function \eqref{eq:Z}{, written for
the $i\neq j$ form of the interaction as in \eqref{eq:H_def},} thus takes the Gaussian
integral form\footnote{{From now on, for the sake of simplicity, we omit the explicit dependence of the
thermodynamic observables on the control parameters.}}
\begin{equation}
\mathcal Z_{N, \bar x, \beta}=\mathcal G\sum_{\{\bm\sigma\}}\int\prod_{\mu\neq\mu_0}\mathcal D\tau_\mu\;
\exp\Big[\frac{\beta N}{2}{\Lambda_\star}\,m^2
+\sqrt{\frac{\beta}{N}}\sum_{\mu\neq\mu_0}\sqrt{\Lambda_\mu}\,\tau_\mu\SOMMA{i=1}{N}\xi^\mu_i\sigma_i
+\frac{\beta\sigma_\eta}{\sqrt N}\sum_{i<j}z_{ij}\sigma_i\sigma_j\Big],
\label{eq:Z_linearized}
\end{equation}
which is the starting point of the interpolation. The prefactor
\begin{equation}
\mathcal G:=\exp\Big(-\frac{\beta}{2N}\sum_{\mu\neq\mu_0}\Lambda_\mu\SOMMA{i=1}{N}(\xi^\mu_i)^2\Big)
\label{eq:diagonal}
\end{equation}
removes the self-interaction terms that the squares generate, restoring the $i\neq j$ form of the original
Hamiltonian \eqref{eq:H_def}.  
The self-interaction of the condensed
pattern is a single term and contributes $O(N^{-1})$, so it is neglected.

\subsection{The interpolating model}
\label{sec:interpolating}

We can now set up Guerra's scheme. As is standard, we introduce a one-parameter family of models connecting
\eqref{eq:Z_linearized}, recovered at $t=1$, to a one-body system, recovered at $t=0$, whose pressure is
computed exactly; the pressure of the former is then obtained from the latter through the fundamental theorem
of calculus,
\begin{equation}
{\mathcal A_{N,\bar x}(1)=\mathcal A_{N,\bar x}(0)+\int_0^1\frac{d\mathcal A_{N,\bar x}}{dt}\,dt} .
\label{eq:FTC}
\end{equation}
Following the physics of the model we allow four scalar parameters $C_1,C_2,C_3,C_4$ in the interpolating
structure, which are left free and fixed a posteriori, once the sum rule is almost achieved. Let
$\bm\theta=\{\theta_i\}_{i=1,\dots,N}$ and $\bm\psi=\{\psi_\mu\}_{\mu\neq\mu_0}$ be standard Gaussian fields,
independent of the quenched disorder and of each other. The interpolating partition function is
\begin{equation}
\mathcal Z_{N, \bar x}(t):=\mathcal G\sum_{\{\bm\sigma\}}\int\prod_{\mu\neq\mu_0}\mathcal D\tau_\mu\;\mathcal B(t),
\label{eq:Zt}
\end{equation}
with the interpolating Boltzmann factor
\begin{equation}
\begin{split}
\mathcal B(t):=\exp\Bigg\{&
\frac{t\beta N}{2}{\Lambda_\star}\,m^2
\;+\;(1-t)\,C_4\,N\,m
\;+\;\sqrt{\frac{t\beta}{N}}\sum_{\mu\neq\mu_0}\sqrt{\Lambda_\mu}\,\tau_\mu\SOMMA{i=1}{N}\xi^\mu_i\sigma_i
\;+\;\sqrt t\,\frac{\beta\sigma_\eta}{\sqrt N}\sum_{i<j}z_{ij}\sigma_i\sigma_j\\[2pt]
&+\;C_1\sqrt{1-t}\SOMMA{i=1}{N}\theta_i\sigma_i
\;+\;C_2\sqrt{1-t}\sum_{\mu\neq\mu_0}\sqrt{\Lambda_\mu}\,\psi_\mu\tau_\mu
\;+\;\frac{1-t}{2}\,C_3\sum_{\mu\neq\mu_0}\Lambda_\mu\,\tau_\mu^2\Bigg\},
\end{split}
\label{eq:Bt}
\end{equation}
the interpolating quenched pressure and the generalized averages being
\begin{equation}
\mathcal A_{N, \bar x}(t):=\frac1N\,\E\big[\ln{\mathcal Z_{N,\bar x}(t)}\big],
\qquad
\omega_t(\,\cdot\,):=\frac{\mathcal G}{{\mathcal Z_{N,\bar x}(t)}}\sum_{\{\bm\sigma\}}\int\prod_{\mu\neq\mu_0}
\mathcal D\tau_\mu\;(\,\cdot\,)\,\mathcal B(t),
\qquad
\langle\,\cdot\,\rangle_t:=\E\big[\omega_t(\,\cdot\,)\big],
\label{eq:At}
\end{equation}
where now $\E$ averages over $\bm\xi$, $\bm z$, $\bm\theta$ and $\bm\psi$.

The two endpoints are the ones we want. At $t=1$ the four terms carrying the parameters vanish, the Gaussian
measure $\mathcal D\tau$ is left untouched and \eqref{eq:Bt} reduces to \eqref{eq:Z_linearized}, so that
${\mathcal A_{N,\bar x}(1)=\mathcal A_{N,\bar x}}$ is the pressure of the original model. At $t=0$ the whole system factorizes
into one-body problems. 

\subsection*{The streaming}
\label{sec:streaming}

Since $\partial_t\mathcal Z_{N,\bar x}/\mathcal Z_{N,\bar x}=\omega_t(\partial_t\mathcal B/\mathcal B)$, the derivative of the
interpolating pressure splits into seven contributions, one per term of \eqref{eq:Bt}. Computing them one at
a time, which is done in Appendix \ref{app:streaming}, one finds the exact identity
\begin{equation}
\begin{split}
\frac{d\mathcal A_{N,\bar x}}{dt}=\ &\frac{\beta{\Lambda_\star}}{2}\big\langle m^2\big\rangle_t-C_4\big\langle m\big\rangle_t
+\frac12\big(\beta-C_2^2-C_3\big)\big\langle p_{11}\big\rangle_t\\[2pt]
&-\frac{\beta}{2}\big\langle p_{12}\,q_{12}\big\rangle_t+\frac{C_2^2}{2}\big\langle p_{12}\big\rangle_t
-\frac{C_1^2}{2}\big(1-\big\langle q_{12}\big\rangle_t\big)
+\frac{\beta^2\sigma_\eta^2}{4}\big(1-\big\langle q_{12}^2\big\rangle_t\big),
\end{split}
\label{eq:streaming_free}
\end{equation}
valid for every $t$ and every $N${, up to terms of order $N^{-1}$,} and for any value of the four parameters.

This is the point at which the parameters are fixed. The quantity $\langle p_{11}\rangle_t$ is not one of the
order parameters of the theory and does not vanish under \eqref{eq:RS}, so the only way to close the sum rule
is to remove it, which forces $C_3=\beta-C_2^2$. The remaining freedom is used to turn each of the surviving
terms into a {centered} fluctuation, and the choice that achieves this is
\begin{equation}
C_1=\sqrt{\beta\bar p+\beta^2\sigma_\eta^2\,\bar q},
\qquad
C_2=\sqrt{\beta\bar q},
\qquad
C_3=\beta(1-\bar q),
\qquad
C_4=\beta{\Lambda_\star}\bar m ,
\label{eq:constants}
\end{equation}
where $C_1$ splits between the two sources of noise, the part $\beta\bar p$ being what the non-condensed
patterns send to a neuron and the part $\beta^2\sigma_\eta^2\bar q$ what the glassy coupling sends. With
\eqref{eq:constants} the streaming becomes
\begin{equation}
\begin{split}
\frac{d\mathcal A_{N, \bar x}}{dt}=\ &
\frac{\beta{\Lambda_\star}}{2}\big\langle(m-\bar m)^2\big\rangle_t
-\frac{\beta}{2}\big\langle(p_{12}-\bar p)(q_{12}-\bar q)\big\rangle_t
-\frac{\beta^2\sigma_\eta^2}{4}\big\langle(q_{12}-\bar q)^2\big\rangle_t\\[2pt]
&-\frac{\beta{\Lambda_\star}}{2}\bar m^2
-\frac{\beta}{2}\,\bar p\,(1-\bar q)
+\frac{\beta^2\sigma_\eta^2}{4}\,(1-\bar q)^2 .
\end{split}
\label{eq:streaming_exact}
\end{equation}
Letting $N\to\infty$ at fixed $\bar x$, and assuming that \eqref{eq:RS} holds also for the generalized
averages $\langle\,\cdot\,\rangle_t$ at every $t\in[0,1]$, with the same values $\bar m$, $\bar q$ and
$\bar p$, the first line of \eqref{eq:streaming_exact} vanishes, the streaming no longer depends on $t$, and
the integral in \eqref{eq:FTC} reduces to the second line; the cutoff is removed only afterwards, in
Section~\ref{sec:RSpressure}.

\subsection*{The one-body term}
\label{sec:onebody}

At $t=0$ the neurons and the analog variables decouple and the partition function factorizes, so that the
Cauchy condition of \eqref{eq:FTC} can be computed exactly. The neurons contribute a single site partition
function averaged over the field $C_1\theta$, each analog variable contributes a one-dimensional Gaussian
integral, and the prefactor \eqref{eq:diagonal} contributes the constant already discussed. Carrying out the
three computations, which is done in Appendix \ref{app:onebody}, and substituting \eqref{eq:constants},
{we find, in the limit $N\to\infty$ at fixed $\bar x$,}
\begin{equation}
\begin{split}
{\mathcal A_{\bar x}(0):=\lim_{N\to\infty}\mathcal A_{N,\bar x}(0)}=\ &\ln2+\E_\theta\Big[\ln\cosh\big(\beta{\Lambda_\star}\bar m+C_1\,\theta\big)\Big]
-\frac{\beta}{2}\int_0^{\bar x}\Lambda(x)\,dx\\[2pt]
&-\frac12\int_0^{\bar x}\Big[\ln\big(1-\beta\Lambda(x)(1-\bar q)\big)
-\frac{\beta\Lambda(x)\,\bar q}{1-\beta\Lambda(x)(1-\bar q)}\Big]dx ,
\end{split}
\label{eq:A0}
\end{equation}
with $\theta\sim\mathcal N(0,1)$. The integral is well defined as long as
\begin{equation}
\beta\,\Lambda(0)\,(1-\bar q)=\beta(1-\bar q)<1 ,
\label{eq:condition}
\end{equation}
since $\Lambda(x)\le\Lambda(0)=1$ for every $x$: the most recent pattern is the channel that saturates first,
exactly as in the Hopfield model.
\subsection{The replica symmetric pressure and the self-consistency equations}
\label{sec:RSpressure}

Inserting \eqref{eq:streaming_exact} and \eqref{eq:A0} into \eqref{eq:FTC} we obtain the replica symmetric
quenched statistical pressure of the Hopfield-Parisi model {at fixed cutoff,
$\mathcal A_{{\rm RS},\bar x}$}, for a target pattern of rescaled age $x_0$ and with
${\Lambda_\star}=\Lambda(x_0)$. {Every integrand decays exponentially on the age scale $x\sim A^2$
by \eqref{eq:Lambda}, so that the error made by truncating the integrals at $\bar x$ is exponentially small in
$\bar x/A^2$, uniformly in the order parameters, as in \eqref{eq:tail}. 
The limit $\bar x\to\infty$ therefore
commutes with the extremization over the order parameters and simply extends the integrals to the whole
positive axis:}\footnote{The integral of $\Lambda$ in the first line of \eqref{eq:ARS} is the constant
$\frac\beta2\int\Lambda=\frac{\beta A^2}{3}$ coming from \eqref{eq:diagonal}, which shifts the pressure but
affects neither the Gibbs measure nor the phase diagram.}
\begin{equation}
\begin{split}
\mathcal A_{\rm RS}(\beta,A,x_0){:=\lim_{\bar x\to\infty}\mathcal A_{{\rm RS},\bar x}}=\ &\ln2
+\E_\theta\Big[\ln\cosh\big(\beta{\Lambda_\star}\bar m+\sqrt{\beta\bar p+\beta^2\sigma_\eta^2\bar q}\;\theta\big)\Big]
-\frac{\beta{\Lambda_\star}}{2}\bar m^2
-\frac{\beta}{2}\int_0^{+\infty}\Lambda(x)\,dx\\[2pt]
&-\frac12\int_0^{+\infty}\Big[\ln\big(1-\beta\Lambda(x)(1-\bar q)\big)
-\frac{\beta\Lambda(x)\bar q}{1-\beta\Lambda(x)(1-\bar q)}\Big]dx\\[2pt]
&-\frac{\beta}{2}\,\bar p\,(1-\bar q)
+\frac{\beta^2\sigma_\eta^2}{4}(1-\bar q)^2 .
\end{split}
\label{eq:ARS}
\end{equation}

The three order parameters are obtained by extremizing \eqref{eq:ARS} with respect to each of them, which is
done in Appendix \ref{app:self}. Writing
\begin{equation}
g:=\beta{\Lambda_\star}\bar m+\sqrt{\beta\bar p+\beta^2\sigma_\eta^2\bar q}\;\theta ,
\label{eq:gfield}
\end{equation}
the derivative with respect to $\bar m$ gives the retrieval equation, the one with respect to $\bar p$ gives
the equation for the spin overlap, and the one with respect to $\bar q$ gives the equation for the overlap of
the analog variables,
\begin{equation}
\bar m=\E_\theta\big[\tanh g\big],
\qquad
\bar q=\E_\theta\big[\tanh^2 g\big],
\qquad
\bar p=\beta\,\bar q\int_0^{+\infty}\frac{\Lambda(x)^2}
{\big[1-\beta\Lambda(x)(1-\bar q)\big]^2}\,dx .
\label{eq:self}
\end{equation}

\begin{remark}
\label{rem:reading}
The age of the target enters \eqref{eq:self} only through $\Lambda_\star=\Lambda(x_0)$, which multiplies the
signal in \eqref{eq:gfield}, whereas the noise is the same for every age. Retrieving an old pattern is
therefore harder not because the interference grows, but because the signal decays.
\end{remark}

\begin{remark}
\label{rem:vanhemmen}
The comparison with the replica theory of nonlinear networks \cite{vanhemmen1987nonlinear}, applied to the
present model in \cite{vanhemmen1988forgetful}, is immediate once the noise is written in the form used
there. Collecting the two contributions to the variance of \eqref{eq:gfield} and setting
\begin{equation}
r:=\frac{\bar p}{\beta}+\sigma_\eta^2\,\bar q ,
\qquad\text{so that}\qquad
\sqrt{\beta\bar p+\beta^2\sigma_\eta^2\bar q}=\beta\sqrt r ,
\label{eq:r}
\end{equation}
the third equation of \eqref{eq:self} gives
\begin{equation}
r=\int_0^{+\infty}\frac{\Lambda(x)^2\,\bar q}{\big[1-\beta\Lambda(x)(1-\bar q)\big]^2}\,dx
\;+\;\sigma_\eta^2\,\bar q ,
\label{eq:r_explicit}
\end{equation}
which is, up to the factor $\bar q$ included in our definition, the noise term of eq.~(8) of
\cite{vanhemmen1988forgetful}, with the sum over the patterns replaced by the integral over the age. The
interpolation and the replica route therefore agree, as they must, at the replica symmetric level, and they
do so at any temperature: \eqref{eq:self} is the system (18)--(20) of \cite{vanhemmen1987nonlinear}
evaluated on the spectrum of the Parisi rule and, for $\sigma_\eta=0$, it coincides with eqs.~(3.8)--(3.9)
of \cite{vanhemmen1987weighted} for a Hopfield model whose weights are the $\Lambda_\mu$. Their solution at
finite temperature is given in Section~\ref{sec:numerics}. Setting $\Lambda(x)\equiv1$ on $(0,\alpha)$ and
$\sigma_\eta=0$ one recovers the Amit-Gutfreund-Sompolinsky solution of the Hopfield model at high storage,
with $r=\alpha\bar q/[1-\beta(1-\bar q)]^2$: consistently with the normalization of \eqref{eq:pab}, our
$\bar p$ and our $r$ are $\alpha$ times the quantities usually denoted by the same letters in that context.
\end{remark}

\begin{remark}
\label{rem:gaussianRS}
The comparison can be pushed one step further, and it provides a test of the Gaussian replacement of
Section~\ref{sec:effective}. Instead of replacing the modes with $|\rho|\ge3$ by the coupling
$\sigma_\eta z_{ij}$, one may keep them in the form \eqref{eq:H_exact}, as further patterns of weight
$\Lambda_\rho$, which is the treatment of \cite{vanhemmen1987nonlinear,vanhemmen1988forgetful}. Each of them
then enters the computation of this section exactly as a non-condensed pattern, and since their weights
vanish one may expand in $\max_{|\rho|\ge3}|\Lambda_\rho|$. To leading order their denominators
$1-\beta\Lambda_\rho(1-\bar q)$ tend to one, so that they contribute to the variance of the field
\eqref{eq:gfield} only their bare squared weight,
$\beta^2\bar q\,N^{-1}\sum_{|\rho|\ge3}\Lambda_\rho^2\to\beta^2\sigma_\eta^2\bar q$, while their
contribution to the pressure reads
\begin{equation}
\frac{\beta^2}{4N}\sum_{|\rho|\ge3}\Lambda_\rho^2\Big[(1-\bar q^{\,2})-2\bar q(1-\bar q)\Big]
\;\longrightarrow\;\frac{\beta^2\sigma_\eta^2}{4}\,(1-\bar q)^2 ,
\label{eq:modes_RS}
\end{equation}
where the first term in brackets collects the one-body term and the corresponding part of the prefactor
\eqref{eq:diagonal}, whose first order contributions cancel, and the second comes from
$-\frac\beta2\bar p(1-\bar q)$. These are exactly the two contributions of the Sherrington-Kirkpatrick
coupling to \eqref{eq:gfield} and to \eqref{eq:ARS}: the glassy term is the limit of a pattern of vanishing
weight, and this is the precise sense in which the modes with $|\rho|\ge3$ behave as a
Sherrington-Kirkpatrick perturbation. The remainder is bounded by a constant times
$\beta^3\sigma_\eta^2\max_{|\rho|\ge3}|\Lambda_\rho|$, and it vanishes in the thermodynamic limit provided
the convergence $\Lambda_\rho\to0$ stated in \cite{vanhemmen1988forgetful} is uniform in $\rho$. At the
replica symmetric level the Gaussian replacement therefore introduces no approximation beyond those of the
replica treatment, in which the residual modes are averaged as independent patterns.
\end{remark}
\begin{figure}[t]
\centering
\includegraphics[width=\linewidth]{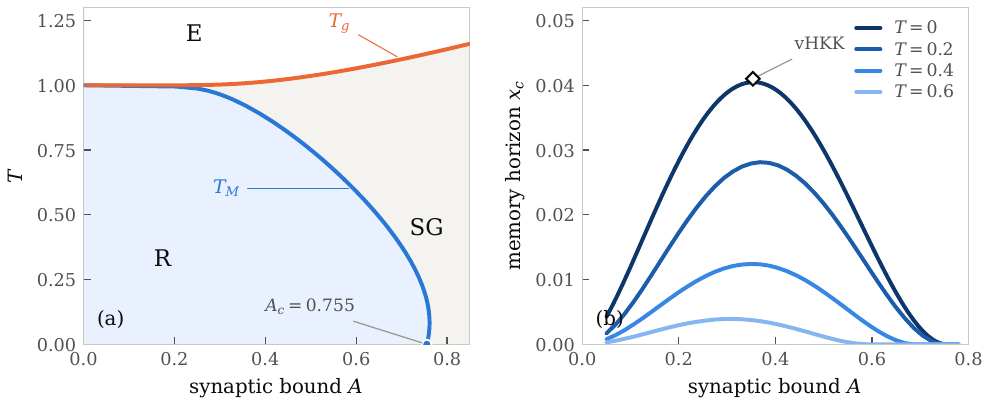}
\caption{Phase diagram of the Hopfield-Parisi model, obtained by solving the
replica symmetric system \eqref{eq:self}.
\emph{(a)} The $(A,T)$ plane for the most recently stored pattern, $x_0=0$ and
hence ${\Lambda_\star}=1$. Three regions are separated. In R the retrieval solution
exists and carries a macroscopic Mattis {magnetization}; in SG the only solution
has $\bar m=0$ and $\bar q>0$, and in E the only solution is $\bar m=\bar q=0$. The line $T_M$ (blue)
is the boundary of the retrieval region, where the solution with $\bar m>0$
ceases to exist. It meets the zero temperature axis at
$A_c=0.755$, marked on the figure, which is the tangency point of
\eqref{eq:self} at $T=0$ and the largest synaptic bound compatible with
retrieval. The line $T_g$ (orange) is the boundary between the ergodic and the
glassy phase, given by {$\beta^2 r/\bar q=1$ in the limit $\bar q\to0$, with $r$ as in
\eqref{eq:r_explicit}}; the transition
across it is continuous.
\emph{(b)} The memory horizon $x_c$ against the bound $A$, at
four temperatures. The dependence on $A$ is not monotone, and the two
mechanisms behind it are visible in \eqref{eq:Lambda} and {\eqref{eq:r_explicit}}: the
weight profile decays on the scale $A^2$, so a small bound gives a short memory
however quiet the network is, while the interference $r$ grows like $A^2$, so a
large bound saturates the synapses with old traces. At $T=0$ the horizon is
maximal at $x_c=0.0405$ for $A=0.351$; the open diamond is the value $0.041$ at
$A=0.354$ reported in \cite{vanhemmen1988forgetful}. Raising the temperature
shrinks the horizon by a factor of ten between $T=0$ and $T=0.6$ while leaving
the optimal bound near $A\simeq0.35$.}
\label{fig:phase}
\end{figure}
\section{Numerical Results}
\label{sec:numerics}

This section collects the solution of the self-consistency equations \eqref{eq:self} and its comparison with
direct simulations of the network. We first solve the system numerically and read from it the phase diagram
and the memory horizon, and we then test both against Monte Carlo runs on networks trained with the rule
\eqref{eq:parisi_rule}. The numerical procedures are described in Appendix \ref{app:numerics}.

\subsection{Numerical solution of the self consistency equations}
\label{sec:numsol}

The system \eqref{eq:self} couples the three order parameters through an integral over the age of the
patterns and a Gaussian average over the field \eqref{eq:gfield}, and we solve it by iteration \footnote{The integral over the age is taken
in the variable $u=x/A^2$, on which the profile collapses onto the single function $\lambda$, so that the
bound enters only as a prefactor; and the boundary of the retrieval region is computed by solving directly
the condition under which \eqref{eq:self} stops admitting a solution with $\bar m>0$. At zero temperature
this is a tangency condition. See Appendix \ref{app:numerics} for further details.}.

Figure \ref{fig:phase}(a) shows the resulting phase diagram in the plane of the control parameters $A$ and $T$, for
the most recently stored pattern. Its shape is the one familiar from the Hopfield model at high load, with
the load replaced by the synaptic bound. Below $T_M$ the system \eqref{eq:self} admits a solution with a
macroscopic Mattis magnetization, and the network retrieves; to the right of it that solution is no longer
available and only the frozen one with $\bar m=0$ and $\bar q>0$ survives, so the network is locked on no
pattern in particular; above $T_g$ even the frozen solution disappears and the only one left is
$\bar m=\bar q=0$. The line $T_M$ meets the zero temperature axis at $A_c=0.755$, and the magnetization the
network still carries there is $\bar m=0.893$, so the retrieval solution does not fade away gradually as the
bound is increased, but disappears while still describing a well recalled pattern. The line $T_M$ is therefore
the limit of existence of the retrieval state;
whether that state is also the one of lowest free energy is a separate question, which we do not address
here. The line $T_g$ never falls below one: the freshest pattern carries weight $\Lambda(0)=1$ by
\eqref{eq:Lambda} and fixes the unit of temperature, and everything the bound adds is further noise, which
can only raise the freezing temperature.

Figure \ref{fig:phase}(b) answers the question the model was built for, namely how far back the network
remembers. We compute the memory horizon, namely,
\begin{equation}
    x_c(A,T):= \sup\{x_0 : \text{Eq.} \, \,\eqref{eq:self} \, \,\text{with} \, \, {\Lambda_\star}=\Lambda(x_0) \, \, \text{admits a solution with} \, \, \bar{m}>0\}.
\end{equation}
The horizon is maximal for an intermediate bound, and the reason is that $A$ enters twice with
opposite signs. It sets the width of the weight profile, since $\Lambda$ decays over ages of order $A^2$, so
a network with rigid synapses forgets quickly no matter how little interference it suffers. It also sets the
strength of that interference, since the noise carries an overall factor $A^2$, so a network with very
plastic synapses keeps so many old traces that none of them is legible. The compromise sits at $A\simeq0.35$,
where the network remembers its last $0.04N$ patterns, and it moves very little with temperature even though
the horizon itself shrinks rapidly.
\begin{remark}
These curves can be compared with the literature at zero temperature, where the replica symmetric equations
were solved for the present model in \cite{vanhemmen1988forgetful}. As noted in Remark \ref{rem:vanhemmen}, our equations coincide with those of \cite{vanhemmen1987nonlinear, vanhemmen1988forgetful}. As a check
of the solver we have applied the same equations to the degenerate spectra of \cite{vanhemmen1987nonlinear},
for which the critical load is reported there, and recovered it to three digits (Appendix
\ref{app:validation}). For the present model we recover the capacity curve of \cite{vanhemmen1988forgetful}
everywhere except close to its end point, with a maximal horizon $x_c=0.0405$ at $A=0.351$ and
$\bar m=0.970$, against $0.041$ at $A=0.354$ with $m=0.97$. The retrieval region closes at $A_c=0.755$,
whereas the threshold quoted in \cite{vanhemmen1988forgetful} corresponds to $A=0.692$ in the present
normalization, and to $A=0.700$ in the simulations of \cite{parisi1986memory}.

The two determinations differ only where $x_c\lesssim0.008$, close to the tangency at which the retrieval
solution disappears by merging with an unstable one, and there the horizon vanishes quadratically,
$x_c\propto(A_c-A)^2$. The last stretch of the curve therefore carries very few patterns, $x_c=0.0032$ at
$A=0.692$ and $0.0025$ at $A=0.700$, and an end point read off as the place where the capacity becomes
indistinguishable from zero depends on the resolution with which a small capacity is detected. Obtaining
$A_c$ as the maximum of an explicit function (Algorithm \ref{alg:fold}) removes this dependence, and the
check against the critical loads of \cite{vanhemmen1987nonlinear}, which are tangency points of the same
kind, tests precisely this step. The value $0.700$ of \cite{parisi1986memory}, with which the threshold of
\cite{vanhemmen1988forgetful} is compared, answers a different question. There a pattern counts as
remembered if the descent started from it ends with no more than $2\%$ of the neurons wrong, that is with an
overlap of at least $0.96$, whereas $A_c$ marks the end of the retrieval state itself, which closes with
$\bar m_c=0.893$. Imposing the same requirement on the zero temperature solution, the most recent pattern
ceases to be recalled at $A=0.724$, and at $A=0.700$ the horizon is already down to $7\times10^{-4}$, less
than one pattern at the sizes $N\le400$ of \cite{parisi1986memory}.
\end{remark}
\begin{figure}[t]
    \centering
    \begin{subfigure}[b]{0.48\linewidth}
        \centering
        \includegraphics[height=5.2cm, width=\linewidth, keepaspectratio]{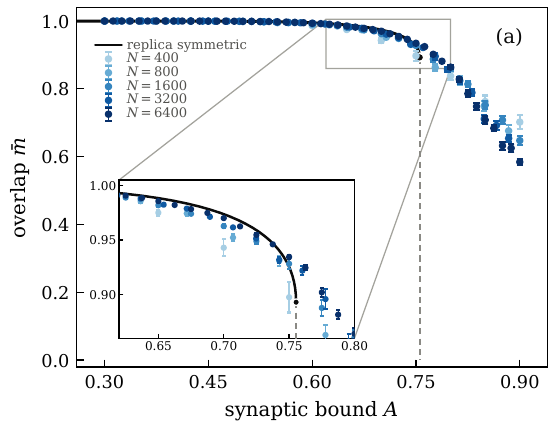}
    \end{subfigure}\hfill
    \begin{subfigure}[b]{0.48\linewidth}
        \centering
        \includegraphics[height=5.2cm, width=\linewidth, keepaspectratio]{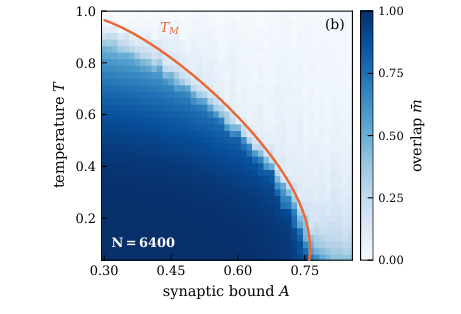}
    \end{subfigure}
    \caption{Retrieval of the most recent pattern in networks trained with the rule \eqref{eq:parisi_rule}.
For each bound a blank network is exposed to $8A^2N+200$ patterns, which places it in the stationary regime
(Section~\ref{sec:hamiltonian}), and the dynamics is started on the most recent pattern with a fraction
$f=0.10$ of the sites flipped. Since $L=A\sqrt N$ has to be an integer, each size is scanned on the finest
grid it admits, of step $1/\sqrt N$. Details are in Appendix \ref{app:numerics}.
\emph{(a)} Zero temperature: overlap reached by an asynchronous descent run to a fixed point, for five sizes
and $50$ realizations of the disorder per bound; error bars are the standard error over realizations. The
solid line is the replica symmetric branch at $T=0$ and ${\Lambda_\star}=1$, which ends at $A_c=0.755$
(dot and dashed line) while still carrying $\bar m_c=0.893$; the inset enlarges the end of the branch.
\emph{(b)} Finite temperature, $N=6400$: overlap with the configuration visited by an asynchronous Metropolis
dynamics, averaged over the last $2\times10^{3}$ sweeps of a run of $4\times10^{3}$ and over $20$ networks
per cell, on a grid of $45$ bounds and $45$ temperatures with step $0.025$ from $T=0.05$; the map is cut at
$T=1$. The line is the boundary $T_M(A)$ of Figure \ref{fig:phase}(a). The freezing line is not shown: the
magnetization vanishes both in the spin glass and in the ergodic phase, so the map carries no information
about it.}
    \label{fig:montecarlo}
\end{figure}

\subsection{Monte Carlo simulations}
\label{sec:montecarlo}

\begin{figure}[t]
    \centering
    \begin{subfigure}[b]{0.48\linewidth}
        \centering
        \includegraphics[height=5.2cm, width=\linewidth, keepaspectratio]{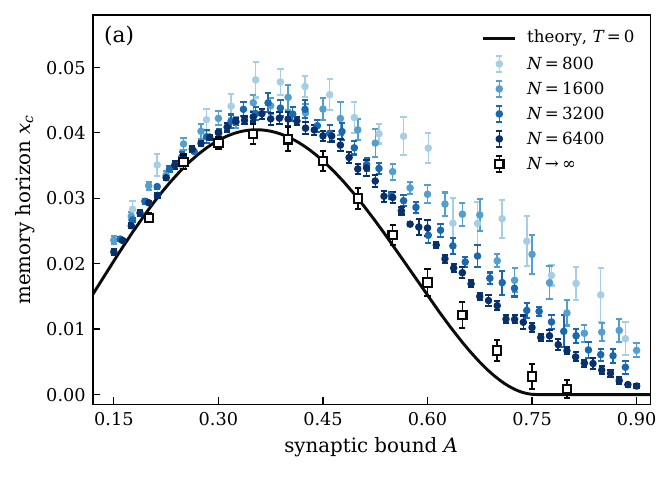}
    \end{subfigure}\hfill
    \begin{subfigure}[b]{0.48\linewidth}
        \centering
        \includegraphics[height=5.2cm, width=\linewidth, keepaspectratio]{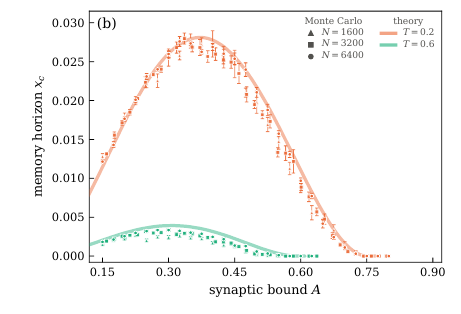}
    \end{subfigure}
    \caption{Memory horizon of networks trained with the rule \eqref{eq:parisi_rule}, prepared and started as
in Figure \ref{fig:montecarlo}. Each network is probed at a sequence of ages of the target pattern, and the
measured horizon is the age at which the overlap, averaged over networks, first falls below $1/2$,
interpolated linearly between neighboring ages; error bars are bootstrap estimates over networks. The
replica symmetric curves, thin in (a) and thick in (b), are those of Figure \ref{fig:phase}(b).
\emph{(a)} Zero temperature, descent to a fixed point, every age from $0$ to $0.1N$ and $10$ networks per
bound. $N=400$ is not shown: there the age is resolved in steps of $1/N=0.0025$, comparable to the horizon
itself near the end of the retrieval region. Open squares: extrapolation to $N\to\infty$ by a fit linear in
$N^{-1/2}$ at each bound, the sizes whose grid does not contain the bound being interpolated in $A$; beyond
$A=0.80$ the measured horizon decays faster than $N^{-1/2}$ and the fit is not shown.
\emph{(b)} $T=0.2$ and $T=0.6$, for $N=1600$, $3200$ and $6400$: Metropolis dynamics with
4000 sweeps, the overlap being averaged over the second half of the run, and 10 networks
per bound.}
    \label{fig:horizon_mc}
\end{figure}

We now test the trained network directly, by starting a dynamics near a stored pattern and recording the
Mattis magnetization reached. Figure~\ref{fig:montecarlo} tests the retrieval region through the overlap of
the most recent pattern, and Figure~\ref{fig:horizon_mc} tests the memory horizon, probing patterns of
every age.

At zero temperature, Figure \ref{fig:montecarlo}(a), the measured overlap follows the replica symmetric
branch and approaches it from below as $N$ grows.  
What the simulations cannot establish is where the branch
ends. The theory predicts that the overlap drops discontinuously at $A_c$, and at finite size this drop is
rounded: beyond $A_c$ the measured overlap decreases only slowly with the bound, and up to $A\simeq0.80$ it
still grows with $N$. The value of $A_c$ therefore rests on the solution of the equations discussed in
Section \ref{sec:numsol}.

At finite temperature, Figure \ref{fig:montecarlo}(b), the region in which the network retrieves follows the
predicted boundary over its whole extent, from $T\simeq0.95$ at small bounds to the vertical descent at
$A_c$, and the color fades across the boundary instead of dropping, which is the rounding expected at finite
size.\footnote{The map starts at $A=0.30$ for two reasons. Below that bound the retrieval region closes
against the freezing line, $T_g-T_M$ being $0.013$ at $A=0.25$ and $0.005$ at $A=0.20$, below the
temperature step of $0.025$, so nothing there is resolvable. And the bound enters the lattice only through
the integer $L=A\sqrt N$, which at $N=6400$ equals $24$ at $A=0.30$, whereas the profile \eqref{eq:Lambda}
is an asymptotic result for large $L$.}

The horizon, Figure \ref{fig:horizon_mc}(a), tells the same story along the bound. Where the memory is long,
the measured horizon is close to the prediction already at the sizes simulated; where it is short, the
measurement exceeds it. This is the excess found beyond $A_c$ in Figure \ref{fig:montecarlo}(a), amplified
near the end of the retrieval region, where the horizon vanishes quadratically and a small shift of the end
point changes it by a large relative amount. The excess shrinks as $N^{-1/2}$, the natural order of finite
size corrections for couplings that move on a lattice of spacing $1/\sqrt N$ (Appendix \ref{app:validation}).
Once extrapolated to $N\to\infty$, the measured horizon agrees with the replica symmetric curve, including
its non monotonic dependence on the bound.

Figure \ref{fig:horizon_mc}(b) extends the test to finite temperature, where the equations had not been
solved before. At $T=0.2$ the measured horizon lies on the prediction already at these sizes, without the
excess found at $T=0$. At $T=0.6$, where the horizon is an order of magnitude shorter, the measured values
fall below the prediction and approach it as $N$ grows. The finite size effect thus changes sign with
temperature, plausibly because at $T=0$ the descent can stop in configurations of large overlap past the end
of the branch, whereas at finite temperature thermal fluctuations can carry the network out of a shallow
retrieval state during the run. 

\section{Conclusions}
\label{sec:conclusions}
We have studied the ``memory which forgets'' introduced by Parisi, a Hopfield network whose synapses are
bounded. Starting from the learning rule, we have shown that it is equivalent to a Hopfield model with
age-weighted patterns superimposed on a Sherrington-Kirkpatrick spin glass, and we have rederived through
Guerra's interpolation its replica symmetric equations, which coincide with those of the replica theory of
nonlinear networks, solving them at finite temperature, where to the best of our knowledge they had not been
solved before. As a pattern ages its signal fades while the noise stays the same, and the twofold role of the
bound produces an optimal value, essentially independent of temperature, at which the memory horizon is
maximal. Monte Carlo simulations of trained networks support these predictions.

Several directions stem from this work. First, our analysis is confined to the replica symmetric level,
which is expected to be only approximate at low temperature, below the de Almeida-Thouless line
\cite{de1978stability}, as discussed in Remark~\ref{rem:RSstatus}. Locating this line, for instance through
the method of \cite{albanese2024almeida}, and moving to broken replica symmetry schemes
\cite{agliari2020replica,albanese2022replica} would sharpen the low temperature portion of the phase
diagram, including the threshold $A_c$ at zero temperature. Second, the network is here exposed to the patterns
themselves, whereas a learning machine experiences noisy examples of archetypes it never sees. Supervised
and unsupervised Hebbian protocols built on examples have been developed for the Hopfield model
\cite{alemanno2023supervised,agliari2022emergence,albanese2024hebbian,ALBANESE2026132009}, and combining them
with bounded synapses raises a question that is absent in both settings separately, namely whether a
network that forgets can still accumulate enough examples of an archetype to infer it, and how the answer
depends on whether a teacher groups the examples or not. Finally, dreaming mechanisms
\cite{crick1983function,hopfield1983unlearning,fachechi2019dreaming,agliari2019dreaming, serricchio2025daydreaming} have been shown to
improve the memorization capabilities of the clipped network \cite{marinari2026dreaming}. The effective
Hamiltonian derived here provides a natural setting for a thermodynamic description of this effect, and in
particular for asking whether dreaming acts by reshaping the weight profile of the patterns, by reducing
the glassy noise, or both.

\acknowledgments
The author is grateful to Andrea Ladiana for useful discussions. \\
The author is a member of the GNFM group within INdAM which is acknowledged. \\
The author acknowledges the use of the Lagrange Multi-GPU Server at the Department of Mathematics, Sapienza University of Rome, for computational resources supporting this work.

\section*{Declaration of generative AI and AI-assisted technologies in the writing process}
During the preparation of this work the author used Claude (Anthropic) to assist in drafting and revising
parts of the text, to improve its language and structure, and to check the internal consistency of the
manuscript. After using this tool, the author reviewed and edited the content as needed and takes full
responsibility for the content of the publication.

\appendix
\section{Detailed computations}
\label{app:details}

This appendix collects the computations leading to \eqref{eq:streaming_free}, \eqref{eq:A0} and
\eqref{eq:self}. Replicas are labeled by superscripts, $\omega_t$ and $\langle\cdot\rangle_t$ are the
generalized averages \eqref{eq:At}, and we use repeatedly the Wick-Isserlis theorem: if $g$ is a standard
Gaussian variable and $F$ is differentiable with moderate growth, then $\E[g\,F(g)]=\E[\partial_gF(g)]$.
Applied to \eqref{eq:At}, for any observable $\mathcal O$ not depending explicitly on $g$,
\begin{equation}
\E\big[g\,\omega_t(\mathcal O)\big]
=\E\Big[\omega_t\Big(\mathcal O\,\frac{\partial_g\mathcal B}{\mathcal B}\Big)
-\omega_t(\mathcal O)\,\omega_t\Big(\frac{\partial_g\mathcal B}{\mathcal B}\Big)\Big],
\label{eq:wick}
\end{equation}
the second term being the one that produces the two replica overlaps, since
$\omega_t(X)\,\omega_t(Y)=\omega_t\big(X^{(1)}Y^{(2)}\big)$.

\subsection{Computation of the streaming}
\label{app:streaming}

Since $\partial_t\mathcal Z_{N, \bar x}/\mathcal Z_{N, \bar x}=\omega_t\big(\partial_t\mathcal B/\mathcal B\big)$,
\begin{equation}
\frac{d\mathcal A_{N, \bar x}}{dt}=\frac1N\,\E\Big[\omega_t\Big(\frac{\partial_t\mathcal B}{\mathcal B}\Big)\Big]
=\sum_{k=1}^{7}\mathcal S_k ,
\label{eq:seven}
\end{equation}
where $\mathcal S_k$ comes from the derivation of the $k$-th term of \eqref{eq:Bt}, in the order in which they
are written there; the prefactor $\mathcal G$ does not depend on $t$ and therefore does not contribute. The
first two terms carry no Gaussian disorder and are immediate,
\begin{equation}
\mathcal S_1=\frac{\beta{\Lambda_\star}}{2}\big\langle m^2\big\rangle_t ,
\qquad
\mathcal S_2=-C_4\big\langle m\big\rangle_t .
\label{eq:S12}
\end{equation}

 Deriving the third term of \eqref{eq:Bt} gives
\begin{equation}
\mathcal S_3=\frac{1}{2N\sqrt t}\sqrt{\frac\beta N}\sum_{\mu\neq\mu_0}\sqrt{\Lambda_\mu}\SOMMA{i=1}{N}
\E\big[\xi^\mu_i\,\omega_t(\tau_\mu\sigma_i)\big].
\end{equation}
The variable $\xi^\mu_i$ appears in $\mathcal B$ only in that same term and at the first power, so that
$\partial_{\xi^\mu_i}\mathcal B/\mathcal B=\sqrt{t\beta\Lambda_\mu/N}\,\tau_\mu\sigma_i$, and \eqref{eq:wick}
with $\mathcal O=\tau_\mu\sigma_i$ gives, using $\sigma_i^2=1$,
\begin{equation}
\E\big[\xi^\mu_i\,\omega_t(\tau_\mu\sigma_i)\big]
=\sqrt{\frac{t\beta\Lambda_\mu}{N}}
\Big(\big\langle\tau_\mu^2\big\rangle_t-\big\langle\tau^{(1)}_\mu\tau^{(2)}_\mu\sigma^{(1)}_i\sigma^{(2)}_i
\big\rangle_t\Big).
\end{equation}
Substituting, the factors $\sqrt t$ cancel and the sum over $i$ produces a factor $N$ and, in the second term,
the overlap $q_{12}$, so that by \eqref{eq:pab}
\begin{equation}
\mathcal S_3=\frac{\beta}{2N}\sum_{\mu\neq\mu_0}\Lambda_\mu
\Big(\big\langle\tau_\mu^2\big\rangle_t-\big\langle\tau^{(1)}_\mu\tau^{(2)}_\mu q_{12}\big\rangle_t\Big)
=\frac{\beta}{2}\big\langle p_{11}\big\rangle_t-\frac{\beta}{2}\big\langle p_{12}q_{12}\big\rangle_t .
\label{eq:S3}
\end{equation}

 With
$\partial_{z_{ij}}\mathcal B/\mathcal B=\sqrt t\,\beta\sigma_\eta/\sqrt N$, the fourth term gives
\begin{equation}
\mathcal S_4=\frac{\beta^2\sigma_\eta^2}{2N^2}\sum_{i<j}
\Big(1-\big\langle\sigma^{(1)}_i\sigma^{(2)}_i\sigma^{(1)}_j\sigma^{(2)}_j\big\rangle_t\Big)
=\frac{\beta^2\sigma_\eta^2}{4}\Big(1-\big\langle q_{12}^2\big\rangle_t\Big)+O(N^{-1}).
\label{eq:S4}
\end{equation}

 With $\partial_{\theta_i}\mathcal B/\mathcal B=C_1\sqrt{1-t}\,\sigma_i$ and
$\partial_{\psi_\mu}\mathcal B/\mathcal B=C_2\sqrt{(1-t)\Lambda_\mu}\,\tau_\mu$, the fifth and sixth terms give
\begin{equation}
\mathcal S_5=-\frac{C_1^2}{2}\big(1-\big\langle q_{12}\big\rangle_t\big),
\qquad
\mathcal S_6=-\frac{C_2^2}{2N}\sum_{\mu\neq\mu_0}\Lambda_\mu
\Big(\big\langle\tau_\mu^2\big\rangle_t-\big\langle\tau^{(1)}_\mu\tau^{(2)}_\mu\big\rangle_t\Big)
=-\frac{C_2^2}{2}\big\langle p_{11}\big\rangle_t+\frac{C_2^2}{2}\big\langle p_{12}\big\rangle_t .
\label{eq:S56}
\end{equation}

 The last one requires no integration by parts,
\begin{equation}
\mathcal S_7=-\frac{C_3}{2N}\sum_{\mu\neq\mu_0}\Lambda_\mu\big\langle\tau_\mu^2\big\rangle_t
=-\frac{C_3}{2}\big\langle p_{11}\big\rangle_t .
\label{eq:S7}
\end{equation}

Summing \eqref{eq:S12}--\eqref{eq:S7} gives \eqref{eq:streaming_free}. Imposing
$C_3=\beta-C_2^2$ removes $\langle p_{11}\rangle_t$, and the choice $C_2^2=\beta\bar q$ turns the two
remaining channel terms into
\begin{equation}
-\frac{\beta}{2}\big\langle p_{12}q_{12}\big\rangle_t+\frac{\beta\bar q}{2}\big\langle p_{12}\big\rangle_t
=-\frac{\beta}{2}\big\langle(p_{12}-\bar p)(q_{12}-\bar q)\big\rangle_t
-\frac{\beta}{2}\bar p\big(\big\langle q_{12}\big\rangle_t-\bar q\big).
\end{equation}
Splitting $C_1^2=\beta\bar p+\beta^2\sigma_\eta^2\bar q$ as in \eqref{eq:constants}, the first piece combines
with the line above,
\begin{equation}
-\frac{\beta}{2}\bar p\big(\big\langle q_{12}\big\rangle_t-\bar q\big)
-\frac{\beta\bar p}{2}\big(1-\big\langle q_{12}\big\rangle_t\big)=-\frac{\beta}{2}\bar p\,(1-\bar q),
\end{equation}
while the second piece combines with \eqref{eq:S4},
\begin{equation}
\frac{\beta^2\sigma_\eta^2}{4}\big(1-\big\langle q_{12}^2\big\rangle_t\big)
-\frac{\beta^2\sigma_\eta^2\bar q}{2}\big(1-\big\langle q_{12}\big\rangle_t\big)
=\frac{\beta^2\sigma_\eta^2}{4}\Big[(1-\bar q)^2-\big\langle(q_{12}-\bar q)^2\big\rangle_t\Big].
\end{equation}
Finally $\tfrac{\beta{\Lambda_\star}}{2}\langle m^2\rangle_t-\beta{\Lambda_\star}\bar m\langle m\rangle_t
=\tfrac{\beta{\Lambda_\star}}{2}\langle(m-\bar m)^2\rangle_t-\tfrac{\beta{\Lambda_\star}}{2}\bar m^2$, which gives
\eqref{eq:streaming_exact}. None of these steps uses \eqref{eq:RS}: the identity holds at finite $N${, up to the
$O(N^{-1})$ remainder of \eqref{eq:S4},} and for every $t$.

\subsection{Computation of the one-body term}
\label{app:onebody}

At $t=0$ the interpolating Boltzmann factor contains no term coupling the neurons to the analog variables, so
\eqref{eq:Zt} factorizes,
\begin{equation}
\mathcal Z_{N, \bar x}(0)=\mathcal G\cdot
\underbrace{\sum_{\{\bm\sigma\}}\exp\Big[\SOMMA{i=1}{N}\big(C_4\,\xi^{\mu_0}_i+C_1\theta_i\big)\sigma_i\Big]}
_{\text{neurons}}
\cdot\underbrace{\prod_{\mu\neq\mu_0}\int\mathcal D\tau_\mu\,
\exp\Big[\frac{C_3\Lambda_\mu}{2}\tau_\mu^2+C_2\sqrt{\Lambda_\mu}\,\psi_\mu\tau_\mu\Big]}_{\text{patterns}} ,
\label{eq:app_Z0}
\end{equation}
where we used $C_4Nm=C_4\sum_i\xi^{\mu_0}_i\sigma_i$.

Now, since $\E[(\xi^\mu_i)^2]=1$, we have:
\begin{equation}
\frac1N\E\big[\ln\mathcal G\big]=-\frac{\beta}{2N}\sum_{\mu\neq\mu_0}\Lambda_\mu
\ \longrightarrow\ -\frac{\beta}{2}\int_0^{\bar x}\Lambda(x)\,dx .
\end{equation}

For the ``neuron'' part of the partition function,  the sum factorizes over the sites and gives
$\prod_i2\cosh(C_4\xi^{\mu_0}_i+C_1\theta_i)$; since $\xi^{\mu_0}_i=\pm1$ and $\theta_i$ is symmetric, the
sign can be absorbed and
\begin{equation}
\frac1N\,\E\Big[\ln\prod_{i=1}^N2\cosh\big(C_4\xi^{\mu_0}_i+C_1\theta_i\big)\Big]
=\ln2+\E_\theta\Big[\ln\cosh\big(C_4+C_1\theta\big)\Big].
\end{equation}

Finally, recalling that $\mathcal D\tau$ carries the weight $e^{-\tau^2/2}$, each factor of
\eqref{eq:app_Z0} is the Gaussian integral
\begin{equation}
\int\frac{d\tau}{\sqrt{2\pi}}\exp\Big[-\frac{a_\mu}{2}\tau^2+b_\mu\tau\Big]
=\frac{1}{\sqrt{a_\mu}}\exp\Big[\frac{b_\mu^2}{2a_\mu}\Big],
\qquad
a_\mu:=1-C_3\Lambda_\mu ,
\quad
b_\mu:=C_2\sqrt{\Lambda_\mu}\,\psi_\mu ,
\label{eq:app_gauss}
\end{equation}
finite precisely under \eqref{eq:condition}. Taking the logarithm and averaging over $\psi_\mu$, whose second
moment is one,
\begin{equation}
\frac1N\sum_{\mu\neq\mu_0}\Big(-\frac12\ln a_\mu+\frac{C_2^2\Lambda_\mu}{2a_\mu}\Big)
\ \longrightarrow\
-\frac12\int_0^{\bar x}\ln\big(1-C_3\Lambda(x)\big)dx
+\frac{C_2^2}{2}\int_0^{\bar x}\frac{\Lambda(x)}{1-C_3\Lambda(x)}\,dx .
\end{equation}
Collecting the three pieces and substituting \eqref{eq:constants} gives \eqref{eq:A0}.

\begin{remark} Equation \eqref{eq:app_gauss} also reproduces the third of \eqref{eq:self}
without any variational argument. At $t=0$ the variable $\tau_\mu$ is Gaussian with mean $b_\mu/a_\mu$ and
variance $1/a_\mu$, and two replicas share the same $\psi_\mu$ but are otherwise independent, so that
$\langle\tau^{(1)}_\mu\tau^{(2)}_\mu\rangle=C_2^2\Lambda_\mu/a_\mu^2$; weighting by $\Lambda_\mu$, summing and
using \eqref{eq:pab} and \eqref{eq:constants},
\begin{equation}
\bar p=\frac1N\sum_{\mu\neq\mu_0}\Lambda_\mu\,\frac{C_2^2\Lambda_\mu}{a_\mu^2}
\ \longrightarrow\ \beta\bar q\int_0^{\bar x}\frac{\Lambda(x)^2}
{\big[1-\beta\Lambda(x)(1-\bar q)\big]^2}\,dx ,
\end{equation}
as it should.
\end{remark}

\subsection{The self-consistency equations}
\label{app:self}

The order parameters are obtained by imposing that the partial derivatives of \eqref{eq:ARS} with respect to
each of them vanish. With $g$ as in \eqref{eq:gfield}, Gaussian integration by parts gives
\begin{equation}
\frac{\partial}{\partial C_1^2}\,\E_\theta\big[\ln\cosh g\big]
=\frac{1}{2C_1}\E_\theta\big[\theta\tanh g\big]
=\frac12\Big(1-\E_\theta\big[\tanh^2g\big]\Big),
\label{eq:app_dlncosh}
\end{equation}
with $C_1^2=\beta\bar p+\beta^2\sigma_\eta^2\bar q$. Then
\begin{equation}
\frac{\partial\mathcal A_{\rm RS}}{\partial\bar m}
=\beta{\Lambda_\star}\Big(\E_\theta\big[\tanh g\big]-\bar m\Big)=0 ,
\end{equation}
which is the first of \eqref{eq:self}, and, since $\partial C_1^2/\partial\bar p=\beta$,
\begin{equation}
\frac{\partial\mathcal A_{\rm RS}}{\partial\bar p}
=\frac{\beta}{2}\Big(1-\E_\theta\big[\tanh^2g\big]\Big)-\frac{\beta}{2}(1-\bar q)=0 ,
\end{equation}
which is the second. Finally, writing ${D}(x):=1-\beta\Lambda(x)(1-\bar q)$, so that
$\partial {D}/\partial\bar q=\beta\Lambda(x)$ and
\begin{equation}
\frac{\partial}{\partial\bar q}\Big[-\frac12\ln {D}+\frac{\beta\Lambda\bar q}{2{D}}\Big]
=-\frac{\beta\Lambda}{2{D}}+\frac{\beta\Lambda}{2{D}}-\frac{\beta^2\bar q\Lambda^2}{2{D}^2}
=-\frac{\beta^2\bar q\,\Lambda^2}{2{D}^2},
\end{equation}
the derivative with respect to $\bar q$ reads
\begin{equation}
\frac{\partial\mathcal A_{\rm RS}}{\partial\bar q}
=\frac{\beta^2\sigma_\eta^2}{2}\Big(1-\E_\theta\big[\tanh^2g\big]\Big)
-\int_0^{+\infty}\frac{\beta^2\bar q\,\Lambda(x)^2}{2\,{D}(x)^2}dx
+\frac{\beta}{2}\,\bar p
-\frac{\beta^2\sigma_\eta^2}{2}(1-\bar q) .
\end{equation}
Once the second of \eqref{eq:self} holds, the first and the last term cancel, and setting the remainder to
zero gives the third of \eqref{eq:self}.

\section{Numerical simulations details}
\label{app:numerics}

This appendix describes the three procedures used in Section \ref{sec:numerics}: the iterative solution of
the self-consistency equations at finite temperature, the determination of the boundary of the retrieval
region at zero temperature, and the Monte Carlo test on the trained network.

\subsection{Solution of the self-consistency equations}

Two quadratures are needed. The integral over the age is taken in the variable $u=x/A^2$, on which
$\Lambda(x)=\lambda(u)$, so that $\int_0^\infty\Lambda(x)^2dx=A^2\int_0^\infty\lambda(u)^2du$ and the bound
appears only as a prefactor; the profile has a square root cusp at the origin, which is removed by the
further substitution $u=s^2$, after which a uniform grid in $s$ is accurate. The resulting values are stable
to four digits between $2\,000$ and $16\,000$ nodes, and the same quadrature returns
$\int\lambda\,du=2/3$ and $\int\lambda^2du=0.28115$, hence $\sigma_\eta^2=0.05218\,A^2$ through
\eqref{eq:sigma}.

The Gaussian average over $\theta$ deserves a comment, because the obvious choice fails. A Gauss-Hermite
rule with a fixed number of nodes produces spurious fixed points at low temperature: there $\tanh g$ is
nearly a step function of $\theta$, and a quadrature whose nodes do not move turns the right hand side of
\eqref{eq:self} into a staircase, whose steps are stable under iteration. Increasing the number of nodes
does not cure the problem. We therefore use a uniform grid on $\theta\in[-10,10]$ with Simpson weights
multiplied by the Gaussian density, which resolves the step wherever it falls.

\begin{algorithm}[t]
\DontPrintSemicolon
\KwIn{bound $A$, inverse temperature $\beta$, weight of the target ${\Lambda_\star}$, tolerance $\epsilon$,
damping $\delta$, threshold $\epsilon_m$}
\KwOut{$(\bar m,\bar q,\bar p)$, or \textsc{none} if no retrieval solution exists}
$\{(u_k,w_k)\}\leftarrow$ quadrature nodes and weights for $\int_0^{u_{\max}}\!\cdot\;du$, from a uniform
grid in $s$ with $u=s^2$\;
$\lambda_k\leftarrow\lambda(u_k)$\;
$\{(\theta_l,{c_l})\}\leftarrow$ Simpson nodes on $[-10,10]$ with weights ${c_l}\propto e^{-\theta_l^2/2}$,
{normalized} to one\;
$\bar m\leftarrow1$,\quad$\bar q\leftarrow0.9$\;
\Repeat{$\max\{|\Delta\bar m|,|\Delta\bar q|\}<\epsilon$}{
  \lIf{$\beta\lambda_k(1-\bar q)\ge1$ for some $k$}{\KwRet \textsc{none}}
  $\bar p\leftarrow\beta\,\bar q\,A^2\sum_k w_k\,\lambda_k^2\big/\big[1-\beta\lambda_k(1-\bar q)\big]^2$\;
  $C\leftarrow\sqrt{\beta\bar p+\beta^2\sigma_\eta^2\bar q}$\;
  $g_l\leftarrow\beta{\Lambda_\star}\bar m+C\,\theta_l$ for every $l$\;
  $\Delta\bar m\leftarrow\sum_l{c_l}\tanh g_l-\bar m$\;
  $\Delta\bar q\leftarrow\sum_l{c_l}\tanh^2g_l-\bar q$\;
  $\bar m\leftarrow\bar m+\delta\,\Delta\bar m$,\quad$\bar q\leftarrow\bar q+\delta\,\Delta\bar q$\;
}
\lIf{$\bar m<\epsilon_m$}{\KwRet \textsc{none}}
\KwRet $(\bar m,\bar q,\bar p)$\;
\caption{Replica symmetric solution of \eqref{eq:self} at given $(A,\beta,{\Lambda_\star})$}
\label{alg:self}
\end{algorithm}

Algorithm \ref{alg:self} implements the iteration. The damping $\delta$ is needed because the map is
contractive only well inside the retrieval region; we use $\delta=1/2$, $\epsilon=10^{-11}$ and
$\epsilon_m=10^{-3}$. The last threshold is what distinguishes a retrieval solution from the frozen one, and
it has to be small: close to $T_g${the retrieval solution appears with a small $\bar m$, which leaves zero smoothly}, so a
generous threshold would cut the retrieval region short. The guard on $\beta\lambda_k(1-\bar q)$ is the
condition under which the integral in the third of \eqref{eq:self} diverges, and it marks the region where no
solution of the assumed form exists. The line $T_M$ of Figure \ref{fig:phase}(a) is obtained by bisecting on
$A$ at fixed temperature until Algorithm \ref{alg:self} stops returning a solution, with a tolerance of
$2\cdot10^{-5}$ on $A$; the line $T_g$ is obtained by solving {$\beta^2r/\bar q=1$ in the limit $\bar q\to0$}, which is an
equation in $\beta$ alone once $A$ is fixed.

\subsection{The boundary of the retrieval region at zero temperature}

At $T=0$ the equations can be reduced and the boundary obtained without iterating. Writing
$Q:=\beta(1-\bar q)$, which stays finite as $\beta\to\infty$, the system \eqref{eq:self} becomes
\begin{equation}
\bar m=\mathrm{erf}\Big(\frac{{\Lambda_\star}\bar m}{\sqrt{2r}}\Big),
\qquad
Q=\sqrt{\frac{2}{\pi r}}\,\exp\Big(-\frac{{\Lambda_\star}^2\bar m^2}{2r}\Big),
\qquad
r=A^2{\mathcal R}(Q),
\label{eq:app_T0}
\end{equation}
with ${\mathcal R}(Q)=\int_0^\infty\lambda(u)^2\big/\big(1-\lambda(u)Q\big)^2du+\sigma_\eta^2/A^2$. Setting
$y:={\Lambda_\star}\bar m/\sqrt{2r}$ turns \eqref{eq:app_T0} into a {parametrization}: every $y$ determines
$\bar m$, then $r$, then $Q$, and finally the bound $A$ at which that triple solves the system. The retrieval
branch is thus traced out as $y$ runs over the positive half line, and the boundary is the value of $A$ at
which the branch turns back, that is the maximum of $A(y)$. This is the tangency condition referred to in
Section \ref{sec:numsol}.

\begin{algorithm}[t]
\DontPrintSemicolon
\KwIn{weight of the target ${\Lambda_\star}$, grid $\{y_j\}$ on $(0,y_{\max}]$, quadrature $\{(u_k,w_k)\}$}
\KwOut{the bound $A_c({\Lambda_\star})$ at which retrieval ceases, and the overlap $\bar m_c$ carried there}
$\mathcal S\leftarrow\emptyset$\;
\ForEach{$y_j$}{
  $\bar m_j\leftarrow\mathrm{erf}(y_j)$\;
  $r_j\leftarrow\big({\Lambda_\star}\bar m_j/(\sqrt2\,y_j)\big)^2$\;
  $Q_j\leftarrow\sqrt{2/(\pi r_j)}\;e^{-y_j^2}$\;
  \lIf{$Q_j\,\max_k\lambda_k\ge1$}{discard $y_j$ and pass to the next}
  ${\mathcal R_j}\leftarrow\sum_kw_k\lambda_k^2\big/\big(1-\lambda_kQ_j\big)^2+\sigma_\eta^2/A^2$\;
  $\mathcal S\leftarrow\mathcal S\cup\{(\sqrt{r_j/{\mathcal R_j}},\,\bar m_j)\}$\;
}
\KwRet the pair of $\mathcal S$ with the largest first entry\;
\caption{Boundary of the retrieval region at $T=0$}
\label{alg:fold}
\end{algorithm}

Algorithm \ref{alg:fold} returns $A_c=0.755$ and $\bar m_c=0.893$ for ${\Lambda_\star}=1$. The memory horizon
follows by inversion: $A_c({\Lambda_\star})$ is increasing, so for a given bound there is a unique weight
${\Lambda_\star}$ at which the branch turns back, and the horizon is the age at which the profile takes that
value, $x_c=A^2\lambda^{-1}({\Lambda_\star})$. At finite temperature the same inversion is performed with
Algorithm \ref{alg:self} in place of Algorithm \ref{alg:fold}.

\subsection{Monte Carlo simulations}

The network is simulated in the variables in which it is defined. A coupling performs a random walk on the
lattice $\{-{L},\dots,{L}\}\varepsilon$ with ${L}=A\sqrt N$, so we store the integer position and divide by
$\sqrt N$ only when the physical coupling is needed, which keeps the whole matrix in one byte per entry for
the sizes used here. The bound is adjusted to $A={L}/\sqrt N$ with ${L}$ an integer, as in {Section~\ref{sec:hamiltonian}}.

Every run starts from a blank network and presents {$K=\lceil\kappa A^2N\rceil+200$} patterns with
{$\kappa=8$}. The choice is dictated by {the relaxation of the walk discussed in Section~\ref{sec:hamiltonian}}: the slowest mode of the reflecting
walk decays as $e^{-\pi^2x/8A^2}$, so at $x=8A^2$ the memory of the initial condition is suppressed by
$e^{-\pi^2}\simeq5\cdot10^{-5}$, and the network is stationary without any tuning of the initial state.

\begin{algorithm}[t]
\DontPrintSemicolon
\KwIn{size $N$, bound $A$, presentation depth {$\kappa$}, corrupted fraction $f$, target ages
$\mathcal M$, temperature $T$, sweep budgets $n_{\rm eq}$ and $n_{\rm av}$ (used only if $T>0$)}
\KwOut{the overlaps $\{\bar m_\mu\}_{\mu\in\mathcal M}$}
${L}\leftarrow\mathrm{round}(A\sqrt N)$,\quad$A\leftarrow {L}/\sqrt N$,\quad
{$K\leftarrow\lceil \kappa A^2N\rceil+200$}\;
$J_{ij}\leftarrow0$ for all $i,j$ \tcp*{integer position of the walk}
\For{$k=1$ \KwTo $K$}{
  draw $\bm\xi^k$ uniform on $\{-1,1\}^N$\;
  $J_{ij}\leftarrow\max\big\{-{L},\min\{J_{ij}+\xi^k_i\xi^k_j,\,{L}\}\big\}$ for all $i\neq j$\;
}
$J_{ii}\leftarrow0$ for all $i$\;
\ForEach{$\mu\in\mathcal M$}{
  $\bm\sigma\leftarrow\bm\xi^\mu$, then flip $\lfloor fN\rfloor$ sites chosen uniformly without
  replacement\;
  \eIf{$T=0$}{
    \Repeat{no site has changed during the sweep}{
      \ForEach{$i$, in a freshly drawn random order}{
        $\sigma_i\leftarrow\mathrm{sign}(h_i)$,\quad $h_i:=\textstyle\sum_jJ_{ij}\sigma_j$\;
      }
    }
    $\bar m_\mu\leftarrow N^{-1}\sum_i\xi^\mu_i\sigma_i$\;
  }{
    \For{$k=1$ \KwTo $n_{\rm eq}+n_{\rm av}$}{
      \ForEach{$i$, in a freshly drawn random order}{
        $\Delta\leftarrow2\sigma_ih_i/N$,\quad $h_i:=\textstyle\sum_jJ_{ij}\sigma_j$\;
        draw {$\zeta$} uniform on $(0,1)$\;
        \lIf{$\Delta\le0$ or ${\zeta}<e^{-\Delta/T}$}{$\sigma_i\leftarrow-\sigma_i$}
      }
      \lIf{$k>n_{\rm eq}$}{$m^{(k)}\leftarrow N^{-1}\sum_i\xi^\mu_i\sigma_i$}
    }
    $\bar m_\mu\leftarrow n_{\rm av}^{-1}\sum_{k>n_{\rm eq}}m^{(k)}$\;
  }
}
\caption{Retrieval test on the trained network, at zero and at finite temperature}
\label{alg:mc}
\end{algorithm}

Both branches of Algorithm \ref{alg:mc} are asynchronous and keep the local field, updating it whenever a
site flips, so that a sweep costs $O(N^2)$ operations regardless of how many sites change. The two differ in
what they read off that field. At $T=0$ only its sign matters and the {normalization} of \eqref{eq:H_def} is
irrelevant, which is why the algorithm works directly in the integer variables. At $T>0$ the {normalization}
has to be restored: in those variables \eqref{eq:H_def} reads $\mathcal H=-N^{-1}\sum_{i<j}J_{ij}
\sigma_i\sigma_j$, so flipping a site costs $\Delta\mathcal H=2\sigma_ih_i/N$, and this is the quantity
entering the acceptance. The field remains an exact integer throughout, so no rounding enters the dynamics
at either temperature.

The initial condition has a fraction $f=0.10$ of the sites flipped, so that its overlap with the pattern
is $0.8$, below the value the retrieval state carries at the boundary, and the measurement asks whether the
pattern has a basin of attraction rather than whether it is a fixed point. At the sizes used the choice
matters little: beyond $A_c$ the descent stops in configurations of large overlap whether it starts from the
pattern or from the corrupted configuration, and at $A=0.90$ and $N=6400$ the overlap reached is $0.581$
from $f=0$ and $0.584$ from $f=0.10$.

The runs of Figure \ref{fig:montecarlo}(a) use $N\in\{400,800,1600,3200,6400\}$ with fifty realizations
of the disorder per bound, the overlap being that of the most recent pattern alone; every descent reaches a
fixed point within $10^3$ sweeps. 
Figure \ref{fig:montecarlo}(b) uses the finite temperature branch at $N=6400$, with
$n_{\rm eq}=n_{\rm av}=2\cdot10^3$ sweeps and $20$ realizations in each cell of a grid of $45$ bounds and
$45$ temperatures, and the same corrupted initial condition, so that the two panels measure the same object
on the two sides of $T=0$. At finite temperature the dynamics has no fixed point, so the stopping rule of the
zero temperature branch is replaced by a fixed budget of sweeps and the overlap is a time average over the
second half of the run; doubling both halves leaves it unchanged within the error of the realizations.

For the horizon, Figure \ref{fig:horizon_mc}, each trained network is probed at a sequence of ages $\mu$ of
the target, so that the overlaps measured at different ages share the same couplings. At $T=0$ the ages are
all those from $0$ to $\lfloor0.1N\rfloor$, with $10$ realizations per bound and $N\in\{800,1600,3200,
6400\}$; at $T=0.2$ and $T=0.6$ the runs use 10 realizations per bound, $N\in\{1600,3200,6400\}$
and $n_{\rm eq}=n_{\rm av}= 2 \cdot 10^3$  sweeps. The measured horizon is the age at which the overlap
averaged over realizations first falls below $1/2$, interpolated linearly between neighboring ages, and its
error is the standard deviation over $400$ bootstrap resamplings of the realizations. At each bound the
extrapolation to $N\to\infty$ is a weighted least squares fit of $x_c(N)=x_\infty+cN^{-1/2}$ to the four
sizes, those whose grid does not contain the bound being interpolated linearly in $A$.

\subsection{Validation of the numerical scheme}
\label{app:validation}

The equations solved here reduce, at $T=0$ and for a degenerate weight profile, to those of
\cite{vanhemmen1987nonlinear}, where the critical load and the overlap carried at the boundary are reported
for two models. Table \ref{tab:validation} compares them with the output of Algorithm \ref{alg:fold} run on
the corresponding spectra.

\begin{table}[h]
\centering
\begin{tabular}{lcccc}
\toprule
 & $\alpha_c$ (here) & $\alpha_c$ \cite{vanhemmen1987nonlinear}
 & $\bar m_c$ (here) & $m_c$ \cite{vanhemmen1987nonlinear}\\
\midrule
Hopfield model   & $0.1379$ & $0.138$ & $0.9674$ & $0.967$\\
clipped synapses & $0.1018$ & $0.102$ & $0.9479$ & $0.948$\\
\bottomrule
\end{tabular}
\caption{Degenerate limit of \eqref{eq:self} at $T=0$, compared with the values reported in
\cite{vanhemmen1987nonlinear}.}
\label{tab:validation}
\end{table}

For the model at hand the value $A_c=0.755$ is stable to four digits when the quadrature is refined from
$2\,000$ to $16\,000$ nodes. The network realizes the profile \eqref{eq:Lambda} only as $L=A\sqrt N\to\infty$.
At finite $L$ the profile is replaced by the exact spectrum of the reflecting random walk on $2L+1$ sites,
the signal of the most recent pattern by $\Lambda_0=2L/(2L+1)$, and the variance of a coupling by
$A(A+\varepsilon)/3$. With these three changes Algorithm \ref{alg:fold} gives $A_c(L)=0.705$, $0.727$,
$0.742$, $0.748$ and $0.752$ for $L=14$, $28$, $60$, $120$ and $240$, which approach the limiting value as
$A_c-0.8/L$. The lattice correction is therefore of order $N^{-1/2}$, the form assumed in the extrapolations
of Section \ref{sec:montecarlo}.
\end{document}